\documentclass[useAMS,usenatbib]{mn2e}

\def\aap{AA}
\def\apjl{ApJL}
\def\mnras{MN}
\def\apj{ApJ}
\def\aj{AJ}

\def\X{{\rm X}}
\def\V{\mathscr{V}}
\def\LL{\mathscr{L}}
\def\G{\mathscr{G}}

\def\s{{s}}
\def\a{{a}}

\usepackage{graphicx}
\usepackage{float}
\usepackage{amssymb}
\usepackage{amsmath} 
\usepackage{deluxetable}
\usepackage{mathrsfs} 

\title[Line Profiles from Discrete Kinematic Data]{Line Profiles from
  Discrete Kinematic Data} \author[N. C. Amorisco and
  N. W. Evans]{N. C. Amorisco$^{1}$\thanks{E-mail:
    amorisco@ast.cam.ac.uk, nwe@ast.cam.ac.uk} and
  N. W. Evans$^{1}$\\ $^{1}$Institute of Astronomy, University of
  Cambridge, Madingley Road, Cambridge CB3 0HA, UK}

\begin{document}

\date{Accepted, Received }

\pagerange{\pageref{firstpage}--\pageref{lastpage}} 

\maketitle

\label{firstpage}

\begin{abstract}
We develop a method to extract the shape information of line profiles
from discrete kinematic data. The Gauss-Hermite expansion, which is
widely used to describe the line of sight velocity distributions
extracted from absorption spectra of elliptical galaxies, is not
readily applicable to samples of discrete stellar velocity
measurements, accompanied by individual measurement errors and
probabilities of membership. These include datasets on the kinematics
of globular clusters and planetary nebulae in the outer parts of
elliptical galaxies, as well as giant stars in the Local Group
galaxies and the stellar populations of the Milky Way. We introduce
two parameter families of probability distributions describing
symmetric and asymmetric distortions of the line profiles from
Gaussianity. These are used as the basis of a maximum likelihood
estimator to quantify the shape of the line profiles. Tests show that
the method outperforms a Gauss-Hermite expansion for discrete data,
with a lower limit for the relative gain of $\approx 2$ for sample
sizes $N \approx 800$. To ensure that our methods can give reliable
descriptions of the shape, we develop an efficient test to assess
the statistical quality of the obtained fit.

As an application, we turn our attention to the discrete velocity
datasets of the dwarf spheroidals (dSphs) of the Milky Way. Sculptor
and Fornax have datasets of $\gtrsim 1000$ line of sight velocities of
probable member stars. In Sculptor, the symmetric deviations are
everywhere consistent with velocity distributions more peaked than
Gaussian. In Fornax, instead, there is an evolution in the symmetric deviations of
the line profile from a peakier to more flat-topped distribution on
moving outwards. Although the datasets for Carina and Sextans are
smaller, they still comprise several hundreds of stars. Our methods
are sensitive enough to detect evidence for velocity distributions
more peaked than Gaussian. These results suggest a radially biased orbital structure
for the outer parts of Sculptor, Carina and Sextans. On the other hand, tangential 
anisotropy is favoured in Fornax. This is all consistent with a
picture in which Fornax may have had a different evolutionary history
to Sculptor, Carina and Sextans.
\end{abstract} 

\begin{keywords}
galaxies: kinematics and dynamics -- Local Group -- galaxies:
individual; Fornax dSph, Sculptor dSph, Carina dSph, Sextans dSph
\end{keywords}
%

%%%%%%%%%%%%%%%%%
\section{Introduction}

Our ability to uncover the elusive properties of dark matter in 
galaxies is based on the analysis of velocities of stars. For the case
of pressure-supported systems like elliptical galaxies, a great
advantage is obtained by considering the properties of the entire line
profile -- that is, the {\it shape} of the line of sight velocity
distributions $\LL(v)$ -- rather than just the first two velocity
moments. This helps break the pernicious mass-anisotropy degeneracy,
which otherwise enables dark matter mass to be traded against
velocity anisotropy at both small and large radii, and hence hidden away.

For elliptical galaxies, higher order velocity information can be
extracted from absorption line spectra. The shape of the velocity
distributions is usually quantified within the framework of a
Gauss-Hermite series, introduced in \citet{Ge93} and \citet{vdMF93}.
Given the velocity distribution $\LL(v)$, the associated Gauss-Hermite series 
is defined by the relation
\begin{equation}
\LL(v)={\gamma
\over{\sqrt{2\pi\sigma^2}}}
{\exp\left[-{1\over2}\left({{v\!-\!\mu}\over
        \sigma}\right)^2\right]}
\left\{1\!+\!\sum_{i=3}^n h_i H_i\left({{v\!-\!\mu}\over \sigma}\right) \right\}
\label{GHdef}
\end{equation}
in which the parameters $\gamma$, $\mu$ and $\sigma$ identify
respectively the normalization, mean and standard deviation of the
best fitting Gaussian, while the coefficients $h_i$ specify the shape
information.  The advantage of this formalism is that Hermite
polynomials $H_i$ are orthonormal with respect to a Gaussian weight
function. The (lowest order) Gauss-Hermite moments measure structure
in the central parts of the velocity distribution and have a limited
dependence on the poorly-determined tails.

However, a disadvantage of the Gauss-Hermite formalism is that it
cannot be easily applied to the large class of problems in which the
kinematic observations come in the form of discrete velocity
measurements, rather than as line of sight velocity distributions
extracted from absorption spectra. This includes the modelling of the
dynamics of galaxies at large radii, where integrated-light
spectroscopy is not possible because of low surface brightness. Here,
tracers such as globular clusters and planetary nebulae are used to
probe the realm of dark matter \citep[e.g.,][]{Ro03, Co09, Na11,
  De12}. In the Local Group, ground-based observations of nearby dwarf
spheroidal galaxies and globular clusters have enabled impressive
datasets of up to thousands of individual line of sight velocities to
be gathered
\citep[e.g.,][]{Kl02,Kl04,Wi04,Ba06,Re06,Ba08,Wa09data,Wa10}. Clusters
of galaxies provide another example in which the kinematic information
is available as discrete velocities \citep[e.g.,][]{Lo03, Wo10}.

For such datasets, the Gauss-Hermite formalism has shortcomings that
limit both accuracy and precision. Difficulties are mainly connected
with the heterogeneous observational uncertainties and the
probabilities of membership. Additionally, a straightforward
implementation of the methods of \citet{Ge93} and \citet{vdMF93} is
only possible for continuous data, thus introducing the potentially
important loss of information due to data binning.  In general, a
dataset of discrete velocities $\vec{V}= \{v_1,\cdots, v_N \}$ comes
together with a set of observational uncertainties
$\vec{\Delta}=\{\delta_1,\cdots ,\delta_N \}$, the values of which are
usually inhomogeneous. Also, the different kinematic tracers may have
different probabilities of membership, $\vec{P}=\{ p_1, \cdots,
p_N\}$, which should also be included.  It is difficult to account
properly for this information within the framework of a Gauss-Hermite
series, as the binning procedure erases virtually everything except
$\vec{V}$. For this reason, the accuracy of the results obtained can
be seriously diminished if the observational uncertainties are large
(in terms of the intrinsic dispersion $\sigma_{\rm int}$), and/or if
either the observational uncertainties themselves or the membership
probabilities are highly inhomogeneous.

Furthermore, whatever the sample size $N$, the shape measured within
the Gauss-Hermite framework is {\it not} the shape of the intrinsic
velocity distribution $\LL$ itself, but rather the shape of its
convolution with the uncertainties' kernel, often taken to be
Gaussian.  This causes an attenuation of the signal due to the
intrinsic deviations from Gaussianity in $\LL$. The magnitude of this
attenuation needs to be separately quantified and then simulated on
models before direct comparison with the observables, which is a
lengthy procedure.

Given these difficulties, it is clear that a feasible solution might
be to use Bayesian methods, which naturally allow us to include all
available information, such as uncertainties of any origin, as well as
probabilities of membership. Unfortunately, the Gauss-Hermite series
is not always positive definite and so does not itself define proper
probability distributions. In order to implement a maximum-likelihood
method, we will have to introduce a new suitable parametrization. At
first glance, this may be viewed as a limitation of a Bayesian
framework, since any parametric family of velocity distributions may
not be flexible enough to describe the dataset. However, we put in
place an analytic device that allows us to test directly whether the
description of the observational sample that is recovered within a
parametric family is a good statistical description or not. Hence, it
is always possible to identify whether the adopted parametrization is
suitable.

As an application for our new methods, we turn our attention to the
highly dark matter dominated dwarf spheroidal galaxies (dSphs) of the
Milky Way. Here, the goal is to map out the density distribution of
the dark matter and compare it to the theoretical predictions of
hierarchical cosmologies. Sometimes, as in the nearby Sculptor dSph,
the properties of the dark halo profile have been strongly constrained
by exploiting the fortunate coexistence of multiple stellar
populations, having different metallicities and kinematics
\citep{Ba08, WaP11, Am12}, whilst \citet{deB12} have mapped out the
detailed star formation history.  \citet{Ja12} have recently shown for
the Fornax dSph, that even with just one (perhaps composite) stellar
population, the detailed modelling of the velocity distributions may
be able to constrain tightly the mass profile. A key ingredient here
is the use of the shape information of the line profiles in addition to the second
moments familiar from straightforward Jeans equation
modelling~\citep[e.g.][]{Gi07,Wa10}.  By themselves, the Jeans
equations do not provide enough information to permit the dark matter
distribution to be mapped out unambiguously~\citep{Ev09}.

We use our new methods to analyze the discrete velocity datasets of
four dSphs -- Sculptor, Carina, Sextans and Fornax -- and obtain for
the first time detailed measurements of the higher velocity
moments. This information provides powerful observables for future
dynamical analyses of the dSphs, and will help constrain the
mass profiles in these systems. Also, since the the formation history
of dSphs is mirrored in their current orbital structure, detailed
information on the line profiles will identify and constrain feasible
formation mechanisms.

The plan of the paper is as follows. In Section~\ref{accuracynotes},
we investigate the magnitude of different effects that influence any
velocity distribution measurement, such as limited sampling,
observational uncertainties and -- for line of sight distributions --
apparent rotation due to systematic proper motion.  In
Section~\ref{ML}, we construct suitable two-parameter families of
distributions to use in a Bayesian likelihood. We describe the method
through which we control the statistical meaningfulness of the maximum
likelihood fit. Section~\ref{appl} deals with the application of the
maximum likelihood measurements of the higher velocity moments to the
dSphs.

%%%%%%%%%%%%%%%%%%%%%%

\section{Achievable Accuracy}\label{accuracynotes}

Reconstructing the intrinsic velocity distribution $\LL(v)$ of a
stellar system is a complex task, since several different effects can
modify the signal that we actually observe. The purpose of this
Section is to quantify these contributions.

\begin{figure}
\centering \includegraphics[width=\columnwidth]{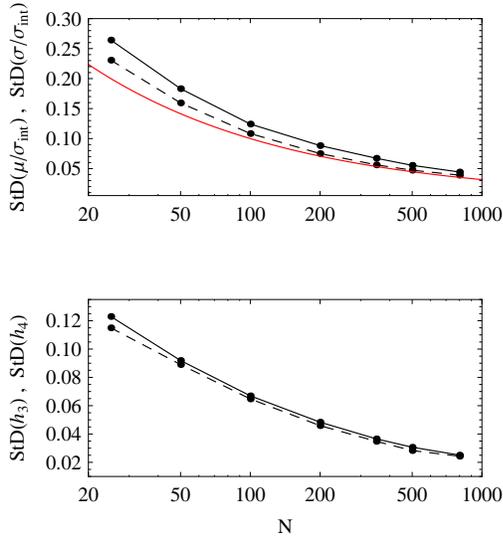}
\caption{Accuracy limits from limited sampling. Upper panel: the
  evolution of the standard deviation of the estimated (normalized) mean $\mu$
  (full line) and dispersion $\sigma$ (dashed line) with the
  sample size $N$. In red, the representative behaviour of $1/\sqrt{N}$
  is shown. Lower panel: the evolution of the standard deviation of
  the Gauss-Hermite moments $h_3$ (full line) and $h_4$ (dashed line)
  with the sample size N.}
\label{nnstarsvdm}
\end{figure}

\subsection{Intrinsic noise from limited sampling}\label{Nnoise}

The most obvious difficulty in measuring the shape of the velocity
distribution in astrophysical systems is limited sampling.  Finite
samples of velocities naturally introduce some noise, which may be
able to alter completely the intrinsic shape.

In general, the magnitude of the noise has its strongest dependence
on the sample size $N$. Hence, the number of available kinematic
tracers poses a limit to the level of achievable accuracy. Some,
smaller dependence can also be ascribed to the method we use to
measure such a shape. In this Section, we use the Gauss-Hermite
expansion introduced in \citet{Ge93} and \citet{vdMF93}. Later, we
will compare these results to our maximum likelihood method.

%Finally, some very small variability in the
%level of achievable accuracy at fixed N can in principle be introduced
%by the details of the shape of $\LL(v)$ itself; however, for our
%present purposes this effect is in practice negligible, and we will
%ignore it for the moment.

As a representative case, we assume that the intrinsic velocity
distribution $\LL$ is a perfect Gaussian, $\G(\mu_{\rm int},
\sigma_{\rm int})$, where $\mu_{\rm int}$ and $\sigma_{\rm int}$ are
its intrinsic mean and dispersion. We generate synthetic samples of
size $N$, namely $\vec{V}= \{v_1,\cdots, v_N \}$, which share the
distribution $\LL(v)=\G(\mu_{\rm int}, \sigma_{\rm int})$. For each of
them, we measure the standard set of properties: $(\mu, \sigma, h_3,
h_4)$, namely the estimated mean, dispersion, and the first non-zero
Gauss-Hermite moments, $h_3$ and $h_4$.

For the sake of clarity, the pair $(\mu, \sigma)$ identifies the best
Gaussian fit $\G(\mu,\sigma)$ to the binned dataset $\vec{V}$ (we do
not report results related to the much less interesting normalization
of the Gaussian fit). The Gauss-Hermite moments are computed as in
\citet{vdMF93}, hence $h_1$ and $h_2$ are identically zero. We adapt
the size of the bin $s_{\rm bin}$ to the size of the sample $N$ with
the standard prescription $s_{\rm bin}\propto N^{-1/3}$.  The bins are
centered on the sample mean, and, as a reference, for our smallest
sample size $N=25$ we use $\approx 4$ bins within the interval
$(-\sigma, \sigma)$.

For any sample size, we quantify the noise due to limited sampling by
repeating this measurement procedure on a large number of synthetic
samples.  Fig.~\ref{nnstarsvdm} shows as a function of sample size
$N$, the variation in the standard deviation (StD) of the estimated
(normalized) mean $\mu/\sigma_{\rm int}$, (normalized) dispersion
$\sigma /\sigma_{\rm int}$, and Gauss-Hermite moments $h_3$ and $h_4$.
As expected, StD$(\mu/\sigma_{\rm int})$ and StD$(\sigma/\sigma_{\rm
  int})$ follow approximately the reference prescription StD$\sim
1/\sqrt{N}$. Note though, that while StD$(\sigma/\sigma_{\rm int})<$
StD$(\mu/\sigma_{\rm int})$, it does not achieve the statistical
prescription StD$(\sigma)\sim 1/\sqrt{2N}$.  More significant,
however, is the result for the Gauss-Hermite moments. We find that, up
to a sample size of $N=200$, the magnitude of the noise (${\rm
  StD}(h_3)\approx {\rm StD}(h_4)\approx 0.05$) is higher than the
amount of intrinsic signal that one, in general, can typically expect
to find in galactic astronomy. This result is approximately
independent of the intrinsic shape of the velocity distribution: we
find that the accuracy limits quantified here remain substantially
unchanged for synthetic datasets extracted from non-Gaussian
distributions.  As a consequence, for sample sizes less than 200, the
accuracy of any measurement is potentially very low, which casts
doubts on the reliability of results obtained using just a few tens of
tracers.

In fact, the situation is even worse, as the intrinsic signal in $\LL$
is also attenuated by the observational uncertainties on the discrete
kinematic measurements. Thus, it is highly likely that any deviation
from Gaussianity detected in small samples is an artefact of
under-sampling and/or binning, rather than being real. We conclude
that it is extremely difficult to measure reliably the shape of any
velocity distribution $\LL$ with a sample size that is significantly
smaller than $N=200$. In the following, we will only consider samples
with $N\geq 200$.

%%%%%%%%
\subsection{Signal attenuation by observational 
uncertainties}\label{attenuation}

\begin{figure}
\centering \includegraphics[width=\columnwidth]{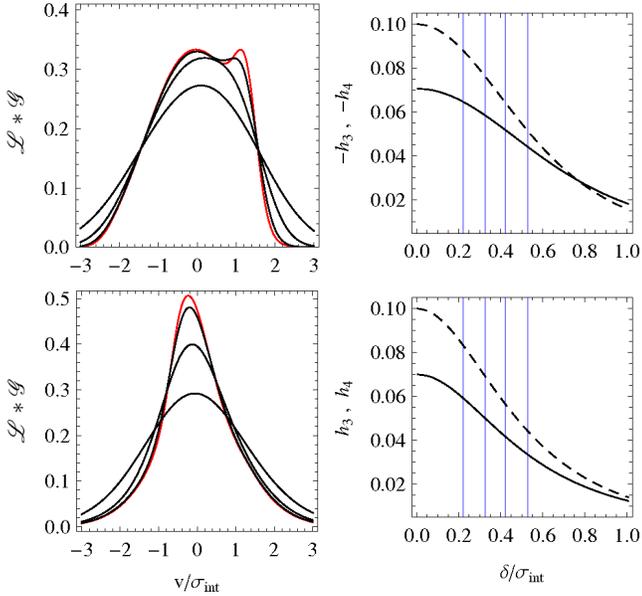}
\caption{The effect of observational uncertainties. Left panels: the
  black full curves show the evolution of $\LL \ast \G$ for
  increasing levels of observational uncertainty. $\LL$ is
  the intrinsic velocity distribution displayed in red. The adopted values
  of $\delta/\sigma_{\rm int}$ are $\{ 0.2, 0.5, 1 \}$.  Right panels: the
  corresponding evolution of the Gauss-Hermite moments $h_3$ (full
  line) and $h_4$ (dashed line). Blue vertical lines illustrate the
  level of observational uncertainty, from left to right, of the
  Fornax, Sculptor, Sextans and Carina datasets
  from~\citet{Wa09data}.}
\label{convsign}
\end{figure}

Inevitably, any real dataset $\vec{V}$ has its own set of
observational uncertainties $\vec{\Delta}=\{\delta_1,\cdots ,\delta_N
\}$. Their effect is to alter the observed velocity distribution, so
that the $i$-th star is in fact associated with the velocity
distribution $\LL \ast \G(0, \delta_i)$, rather than with the
intrinsic $\LL$ itself. By $\LL \ast \G$, we indicate the convolution
of the velocity distribution $\LL$ with the Gaussian kernel $\G$:
\begin{equation}
\left[\LL \ast \G \right](v)=
\int_{-\infty}^{\infty}
dx\ \LL(x)\ {{\exp{\left[-{1\over 2}
        \left({{v-x}\over\delta_i}\right)^2\right]}}\over{\sqrt{2\pi\delta_i^2}}}
\label{convolution}
\end{equation}
Unsurprisingly, the effect of this convolution is to attenuate the
features of $\LL$.

For a given velocity distribution $\LL(v)$, the magnitude of this
attenuation is a function of the dimensionless ratio observational
uncertainty to the intrinsic dispersion $\delta/\sigma_{\rm int}$
only. For any sample size $N$, it is impossible to resolve the
intrinsic velocity distribution $\LL$ with the Gauss-Hermite method
used in previous Section~\ref{Nnoise}. Rather, the measured signal is
the one corresponding to $\LL \ast \G(0, \delta_m)$, where $\delta_m$
is, approximately, the mean of the sample of uncertainties
$\vec\Delta$.

As an example, let us consider two arbitrary velocity distributions
$\LL(v)$ both sharing the same amount of non-Gaussianity as
characterized by their Gauss-Hermite expansion: the first has $(h_3,
h_4)=(-0.07, -0.1)$, and the second has $(h_3, h_4)=(0.07, 0.1)$. They
are displayed as red full curves in the upper and lower left panels of
Fig.~\ref{convsign}. The right panels show the evolution of the
Gauss-Hermite moments $h_3$ (full line) and $h_4$ (dashed line) as
functions of observational uncertainty on the kinematic measurements
$\delta/\sigma_{\rm int}$. Clearly, these are monotonic functions that
tend to zero for both Gauss-Hermite moments -- that is to a perfect
Gaussianity -- when $\delta/\sigma_{\rm int}\gtrsim 1$ or when the
uncertainty is high enough to overwhelm any signal in $\LL$. The black
curves in the left panels illustrate this process by showing $\LL \ast
\G$ when $\delta/\sigma_{\rm int}\in \{0.2, 0.5, 1\}$.

From Fig.~\ref{convsign}, it is clear that the effect of attenuation
is significant. The vertical lines in the right panels show the
levels of average observational uncertainty $\delta_m/\sigma_{\rm int}$
for the datasets on the Galactic dwarf spheroidals presented
by~\citet{Wa09data}.  Specifically, from the lowest to highest
levels of uncertainty, we find
\begin{eqnarray}
{\rm Fornax :}	& \delta_m/\sigma_{\rm int}\approx 0.22 \nonumber\\
{\rm Sculptor :}	& \delta_m/\sigma_{\rm int}\approx 0.33\nonumber\\
{\rm Sextans :}	& \delta_m/\sigma_{\rm int}\approx 0.42\nonumber\\
{\rm Carina :}	& \delta_m/\sigma_{\rm int}\approx 0.53\nonumber
\label{errlevels}
\end{eqnarray} 
Taking the case of Sculptor as an example and consulting
Fig.~\ref{nnstarsvdm}, it is clear that even the strong signal adopted
here ($|h_4|=0.1$) would remain smaller than (or comparable with) the
noise from limited sampling up to $N\approx 100$. On the other hand,
by reversing the argument, if the typical ultrafaint has a kinematic
sample of $N\lessapprox 100$, any signal would be overwhlemed by the
shot noise unless $\delta_m/\sigma_{\rm int}\lessapprox 0.25$.
Although important improvement has been achieved \citep[see for
  example][]{Ko11}, significantly larger datasets would be necessary
to resolve the line profile information in such systems. Finally, let
us note that if we compare the observed shape of $\LL \ast \G$ with
that of theoretical dynamical models, we must account for this
inevitable attenuation effect. Whether using a Gauss-Hermite expansion
or a non-parametric reconstruction of the velocity distribution
$\LL(v)$ -- as in \citet{Ja12} -- this attenuation must be applied to
the dynamical models themselves before comparison with the
observables. On the other hand, a maximum likelihood approach allows
us to take into account the observational uncertainties while deriving
our observables, and thus we are able to reconstruct the properties of
$\LL$ itself, rather than those of $\LL \ast \G$.

%%%%%%%%
\subsection{Apparent rotation from global proper motion}\label{approtsect}

\begin{figure}
\centering \includegraphics[width=\columnwidth]{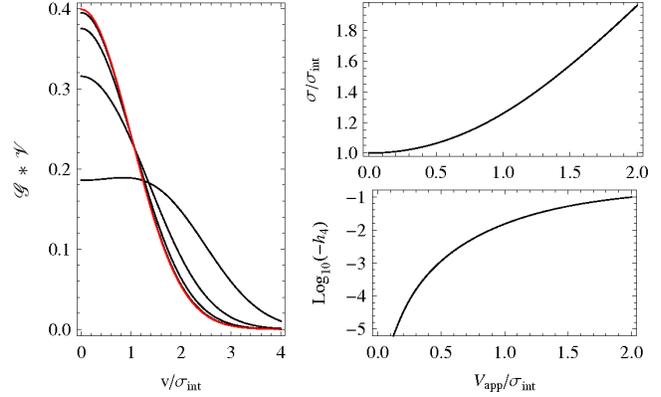}
\caption{The effect of apparent rotation. Left panel: an intrinsic
  Gaussian velocity distribution $\LL=\G(0,\sigma_{\rm int})$ (red curve)
  disturbed by apparent rotation of growing amplitude
  $V_{\rm app}/\sigma_{\rm int}\in\{0.2, 0.5, 1, 2 \}$ (black curves).  Right
  panels: the evolution of the dispersion $\sigma$ of the best-fitting
  Gaussian and of the first nonzero Gauss-Hermite moment $h_4$ for
  different $V_{\rm app}/\sigma_{\rm int}$.}
\label{theorot}
\end{figure}

The exploitation of the projection effect that causes an extended
object in the sky to have an apparent line-of-sight solid-body rotation
as a consequence of its global proper motion has a long
history~\citep[e.g.,][]{Fe61}. Very recently, the effect has been
exploited by \citet{Wa08pm} to derive the systemic proper motion of
the Fornax, Sculptor Carina and Sextans dSphs.

For our purposes, apparent rotation is a potentially dangerous effect
because it alters the shape of the observed velocity distribution. In
particular, for the line of sight velocity distribution $\LL$, the
contribution of apparent rotation is degenerate with the signal
produced by tangential anisotropy.  For this reason, whenever
possible, apparent rotation is usually subtracted from the dataset
before estimating the shape of $\LL$. Nonetheless, it is useful to have
an understanding of this spurious effect, since the subtraction of the
apparent velocity field from the dataset does come at a price.  It
may, in fact, not be worthwhile degrading each discrete velocity
measurement with the uncertainty of the apparent velocity field if the effect
on $\LL$ is expected to be exceedingly small.

For simplicity, we consider the case in which the sample of kinematic
tracers is uniformly distributed on a circle.  This assumption mimics
the more realistic case of a thin circular annulus.  Let us use the
notation
\begin{equation}
v_{\rm app}(R,\theta)=c\ D\ \vec{\mu}\cdot\vec{R} =V_{\rm
  app}\ \sin(\theta-\theta_{\rm app})
\label{approt}
\end{equation}
where $\vec{\mu}$ is the proper motion, $D$ is the distance, $\vec{R}$
is the projected distance on the sky (measured in arcmin),
$\theta_{\rm app}$ identifies the apparent rotation axis, and $c$ is a
constant ($c=1.379\cdot 10^{-5}$ (km century)/(s kpc arcmin mas)). Hence,
$V_{\rm app}$ represents the maximum apparent rotation velocity
attainable at the projected radius $R$. If $\LL$ is the intrinsic
velocity distribution of the tracers on the circle with radius $R$,
then the effective velocity distribution we observe is $\LL \ast
\V(V_{\rm app})$, where
\begin{equation}
\V(V_{\rm app})= 
\begin{cases}{\left(\pi V_{\rm app}\sqrt{1-v^2/V^{2}_{\rm app}}\right)^{-1}} 
           & {\rm if}\ |{v / V_{\rm app}}|\leq1\\ 0 & {\rm if}\ |{v /
    V_{\rm app}}|>1
\end{cases}
\label{vdrot}
\end{equation}
is the velocity distribution associated with the apparent
rotation.

The magnitude of the deviations from $\LL$ introduced by the
convolution is a function of the ratio $V_{\rm app}/\sigma_{\rm int}$
only.  Fig.~\ref{theorot} illustrates the representative case in which
$\LL=\G(0,\sigma_{\rm int})$. The left panel illustrates four different
velocity distributions $\G \ast \V(V_{\rm app})$, obtained for $\V_{\rm
  app}/\sigma_{\rm int}\in\{0.2, 0.5, 1, 2 \}$.  As anticipated,
deviations from the pure Gaussian case (in red) are towards a flat
topped velocity distribution and even a double peaked structure
appears at higher levels of apparent rotation. Both these features are
usually considered indicative of tangential velocity anisotropy,
though clearly this is not the case here.  The panels on the right
quantify the effect of apparent rotation on the dispersion of the best
fitting Gaussian $\sigma$ and on the Gauss-Hermite moment $h_4$. Since
both $\G$ and $\V$ are even functions, $h_3$ is identically zero.
Both $\sigma$ and $h_4$ are more strongly modified as the magnitude of
the apparent rotation increases.

For the dSphs considered later in this paper, the effect is actually
very small. Sculptor and Fornax are the only dSphs where a reliable
non-zero estimate of the rotation signal can presently be obtained.
Even at large distances $R$, the amount of apparent rotation $V_{\rm
  app}$ (less than a few kms$^{-1}$) still remains just a fraction of
the intrinsic dispersion $\sigma_{\rm int}$ (typically $\sim 10$
kms$^{-1}$).  Unless we have a very large sample size, the
modifications introduced by apparent rotation are smaller than the
achievable accuracy. As Fig.~\ref{theorot} shows, this is
particularly true for the Gauss-Hermite moment $h_4$, which remains
virtually unaffected up to $V_{\rm app}/\sigma_{\rm int}\approx 0.7$,
even with sample sizes as large as $N=800$.

This is clearly not a general rule, and Fig.~\ref{theorot} can be used
to assess other cases, such as nearby globular clusters. Also, there
may be reasons that justify the subtraction of the apparent rotation
from the kinematic dataset, as for example if the considered annuli
are highly non-uniformly populated, or if other spatial regions are
considered in the place of annuli, or if the estimate of the apparent
velocity field is precise enough.

%%%%%%%%%%%%%%%%%%%%%%

\section{Maximum likelihood method}\label{ML}

\subsection{Introduction}

Suppose we have a set of $N$ kinematic tracers with velocities
$\vec{V}= \{v_1,\cdots, v_N \}$, which are aligned along the same
axis, for example, the line of sight direction. We assume that they
sample a specified spatial region, the velocity distribution of which
we want to determine. The set $\vec{V}$ is accompanied by the set of
velocity uncertainties $\vec{\Delta}= \{\delta_1,\cdots, \delta_N \}$,
and membership probabilities $\vec{P}=\{ p_1, \cdots, p_N\}$.  We take
these probabilities as assigned constants, although in some cases the
probability $p_i$ may be modelled as a function of the velocity $v_i$
as well as of other observable quantities in order to identify
foreground objects, separate stellar populations and so on.

Now suppose we have at our disposal a family of velocity distributions
$\LL(v)$. This is associated with a set of parameters
$\vec\Theta=\{\theta_1, \cdots, \theta_j\}$.  Within this family, we
can recover the best statistical description of the sample $\vec{V}$
by maximizing the likelihood:
\begin{equation}
L(\vec\Theta)=\prod_{i=1}^{N}\ p_i\ \left[\LL(\vec\Theta) \ast \G(0, \delta_i)\right](v_i)\ ,
\label{lik}
\end{equation}
in which we have implicitly assumed that the distributions $\LL$ have
unit integral.

In the case of a Gauss-Hermite expansion, the set $\vec\Theta$
comprises the dimensional pair $(\mu, \sigma)$ of the best Gaussian
fit, together with the series of dimensionless moments $h_j$,
truncated according to the size of the dataset as well as to the
uncertainties of the kinematic measurements. Note that in the
terminology of \cite{vdMF93}, $\mu$ and $\sigma$ represents the mean
and dispersion of the best-fitting Gaussian.

In this paper, however, we prefer to use $\mu$ and $\sigma$ to denote
the first and second moment of the distribution $\LL$:
\begin{eqnarray}
\mu 		& = & \int_{-\infty}^{\infty}\ \LL(\vec\Theta; v)\ v\ dv\ ;\label{mu}\\
\sigma^2 	& = & \int_{-\infty}^{\infty}\ \LL(\vec\Theta; v)\ v^2\ dv\ .\label{sigma}
\end{eqnarray}
We can highlight the role of these two dimensional quantities in the
likelihood (\ref{lik}) by separating them from the remaining {\it
  shape parameters} in $\vec{\Theta}$, which we group in the subset
$\vec{\Theta}_{\rm sh}$, to yield
\begin{equation}
L(\vec\Theta)=\prod_{i=1}^{N}\ {p_i\over\sigma}\ \left[\LL(\vec{\Theta}_{\rm
    sh}) \ast \G(0,
  \delta_i)\right]\left({{v_i-\mu}\over\sigma}\right)\ .
\label{lik1}
\end{equation}
We have made explicit use of the fact that the distributions
$\LL(\vec{\Theta}_{\rm sh}; v)$ have zero mean, unit integral and unit
dispersion. Finally, we will use the notation $e_{\theta}$ for the
uncertainty of the parameter $\theta\in\vec{\Theta}$. This
uncertainty is defined by the 68\% confidence region associated with
the marginalized likelihood.

The implementation of a maximum likelihood method for measuring the
shape of velocity distributions $\LL$ brings about a number of
advantages. The most evident one is the elimination of any arbitrary
aspect introduced by binning in velocity space. Of equal importance is
the problem of observational uncertainties.  These are not easily --
nor usually -- taken into account by the binning procedure, hence
making any measurement questionable. The method of maximum likelihood
instead furnishes a natural way to incorporate any uncertainty into
the measurement procedure. Also, we can reconstruct directly the
intrinsic velocity distribution $\LL$, rather than the attenuated one
$\LL\ast\G(0,\delta_m)$. This has two consequences. First, the limit
in accuracy due to sampling (see Sect.~\ref{Nnoise}) is less
important, since intrinsic signals are stronger.  Second, observables
obtained in this way can be directly compared with dynamical
models. This is not possible in general, since if
$\LL\ast\G(0,\delta_m)$ is reconstructed, this should be compared with
an analogous quantity which is only indirectly provided by the models.

Given these advantages, it is natural to look for an implementation of
the maximum likelihood approach using the standard Gauss-Hermite
expansion. Such an approach has been proposed in \citet{vdV06} and 
used in \citet{Ra09} to characterize the kinematics of the Galactic bar.
Particular attention must be devoted to the tails of the velocity distributions, as a truncated Gauss-Hermite
series is not always positive definite and so cannot be used as a
probability distribution. In order to circumvent these difficulties, we construct alternative families of
probability distributions $\LL(\vec{\Theta}_{\rm sh}; v)$, and it is
to this problem we now turn.

%%%%%%%%
\subsection{A Two-Parameter Family of Velocity Distributions}\label{family}

For velocity distributions $\LL$, it is useful to start with a
Gaussian profile $\G$, since it represents a good approximation for
most realistic cases.  We maintain this perspective, but adopt two
additional parameters to measure the deviations from a pure Gaussian
profile.  The parameter $s$ quantifies the magnitude of the {\it
  symmetric} deviations, the parameter $a$ the magnitude of {\it
  asymmetric} deviations. This is in addition to the parameters $\mu$
and $\sigma$, which are the mean and the dispersion of any
distribution $\LL$. For the sake of clarity:
\begin{equation}
\vec{\Theta}=\{\mu, \sigma\}\cup \vec{\Theta}_{\rm sh}=\{\mu,\sigma,
\s, \a\}\ .
\end{equation}
As in any parametric approach, the choice of the adopted model is a
crucial step. In the next subsections, we propose families of velocity
distributions encompassing a wide range of deviations from the
Gaussian profile seen in typical stellar dynamical systems.

\begin{figure}
\centering
\includegraphics[width=\columnwidth]{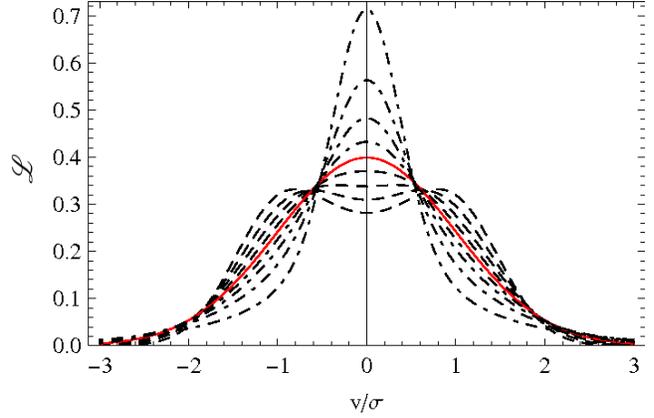}
\caption{Our one-parameter family of symmetric velocity
  distributions. The red profile illustrates a perfect Gaussian;
  dashed profiles display negative values of $s$, ranging in the
  interval $[-4,0)$; dotdashed profiles display positive values of $s$
    ranging in the interval $(0,0.8]$.}
\label{symmdev}
\end{figure}
\begin{figure}
\centering
\includegraphics[width=\columnwidth]{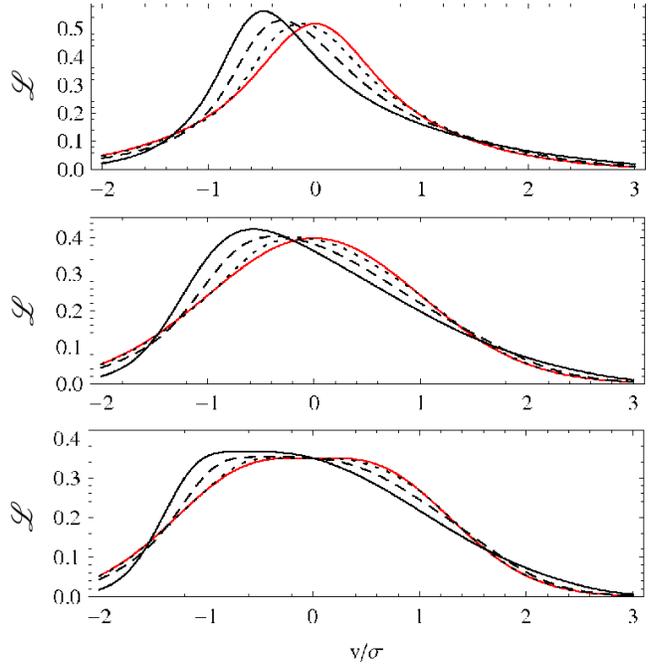}
\caption{An illustration of our final parametrization of asymmetric
  deviations.  In each of the three panels, which display the cases $s
  \in\{0.5, 0, -0.5\}$, the red profile illustrates the symmetric
  distribution $\LL(s,a=0; v)$; black profiles display the growth of
  asymmetries as measured by our parameter $a$, for $a \in\{0.1, 0.3,
  0.7\}$.}
\label{asymmexamples}
\end{figure}

\subsubsection{Symmetric deviations}\label{symm}

In order to build a set of representative, symmetric, non-Gaussian
velocity distributions, we exploit the three-dimensional distribution
of velocities
\begin{equation}
f(v_r, |\vec{v}_t|)={{ |\vec{v}_t|^{-2\s}}\over{\rho(\s)}}\exp\left[-{{v^2_r+|\vec{v}_t|^2}\over {2\sigma^2_r}}\right]\ ,
\label{fconstani}
\end{equation}  
in which $v_r$ and $\vec{v}_t$ are respectively the radial and
tangential components of the velocity, $s$ is our free parameter for
symmetric deviations ($\s=0$ identifies the Gaussian case) and
$\rho(\s)$ is defined so that $\int f d^3\vec{v}=1$:
\begin{equation}
\rho(\s) =  \sqrt{2\pi}\sigma_r^3\ (2 \sigma_r^2 )^{-\s}\ \Gamma(1-\s)\ .
\label{rhoconstani}
\end{equation}  
The simple model of eqn~(\ref{fconstani}) is familiar from constant
anisotropy models $\beta=\s$, where, with the usual notation, $\beta=
1-\sigma^2_t/2\sigma^2_r$. In particular, these three-dimensional or
intrinsic velocity distributions are the constant anisotropy phase
space distribution functions for the isothermal sphere~\citep[see
  e.g.,][]{Ge91, Ev94}. This seems a natural starting point for
galaxies with flattish velocity dispersion profiles, for which we
might plausibly expect the intrinsic velocity distributions to be
reasonably similar.

Our family of symmetric velocity distributions corresponds to a set of
line of sight velocity distributions generated by $f$.  The direction
associated with the line of sight identifies the velocities
$v_{\parallel}$ and $\vec{v}_{\perp}$. If $\varphi$ is the angle
defined by the line of sight and its projection onto the plane of the
tangential velocity $\vec{v}_t$, we have that
\begin{equation}
\left\{
\begin{array}{lclcl}
v_r & = & |v_{\perp}|\cos(\alpha)\cos(\varphi) & + & v_{\parallel}\sin(\varphi)\\
v_{\theta} & = & |v_{\perp}|\cos(\alpha)\sin(\varphi) & - & v_{\parallel}\cos(\varphi)\\
v_{\phi} & = & |v_{\perp}|\sin(\alpha) &&
\end{array}
\right.\ .
\label{vtransf}
\end{equation}
This set of linear transformations allows us to compute the line of
sight distribution generated by $f$ for any direction $\varphi$:
\begin{eqnarray}
f_{los}(v_{\parallel}, \varphi) & = & \int d^2\vec{v}_{\perp}\ f\left[v_r(v_{\parallel}, \vec{v}_{\perp}), v_t(v_{\parallel}, \vec{v}_{\perp})\right]\ ,
\label{losint}\\
 & = & \int_{0}^{\infty} dv_{\perp} \int_{0}^{2\pi}d\alpha\ v_{\perp} f(v_r, |v_t|)\ .\nonumber
\end{eqnarray}
Given the properties of the pressure tensor of $f$, the line of sight
distribution $f_{\rm los}$ has a dispersion
\begin{equation}
\sigma_{\parallel}(\varphi) = \sigma_r\sqrt{1-\s \cos(\varphi)^2}.
\label{constanir}
\end{equation}
Hence, the distributions defined by
\begin{equation}
\LL(\s, \varphi; v)=\sigma_{\parallel}(\varphi)\cdot
f_{los}[\sigma_{\parallel}(\varphi)\cdot v]
\label{unitdisp}
\end{equation}
have by construction zero mean, unit integral and unit dispersion, and
are well-suited for our maximum likelihood method.

By reporting explicitly all functional dependences in
eqn~(\ref{unitdisp}), we highlight the fact that the distributions
$\LL$ have two parameters, namely the shape $s$ and the angle
$\varphi$. The parameter $s$ is associated with genuine deviations
from the Gaussian profile, while the effect of varying $\varphi$
between 0 and $\pi/2$ at fixed $s$ is to erase these deviations
($\varphi=\pi/2$ identifies the radial direction, whose line of sight
distribution is Gaussian for any $s$). For this reason, $s$ and
$\varphi$ cannot be maintained as independent parameters -- since they
are strongly correlated -- and a prescription of the form
$\varphi=\varphi(\s)$ is needed as a `closure'. Different
prescriptions introduce small differences in the resulting family of
distributions, but in this paper we adopt
\begin{equation}
\cos[\varphi(\s)] \equiv 15/(16-\s)\ 
\label{closure}
\end{equation}
for two different reasons. First, for positive $s$, this allows us to
keep non Gaussianities as strong as possible when $s$ approaches its
upper limit of unity $\varphi(s= 1)= 0$. At the same time, we avoid
setting $\varphi$ uniformly to zero, since this produces distributions
with strongly pronounced peaks also for much smaller values of
$s$. Second, the closure condition (\ref{closure}) allows us, for
negative values of $s$, to provide a range of flat topped
distributions before a double peaked structure appears. Flat topped
distributions are almost absent if $\varphi$ is uniformly set to
zero. Fig.~\ref{symmdev} illustrates the family of symmetric
distributions $\LL$ we have defined here.  Both the characteristic
extremes of a spiky distribution with substantial tails and of a
double peaked structure with sharp edges can be clearly identified
within the displayed range $-4\leq s\leq 1$. Between such extremes,
the entire range of intermediate configurations is accessible as well.

\begin{figure}
\centering
\includegraphics[width=\columnwidth]{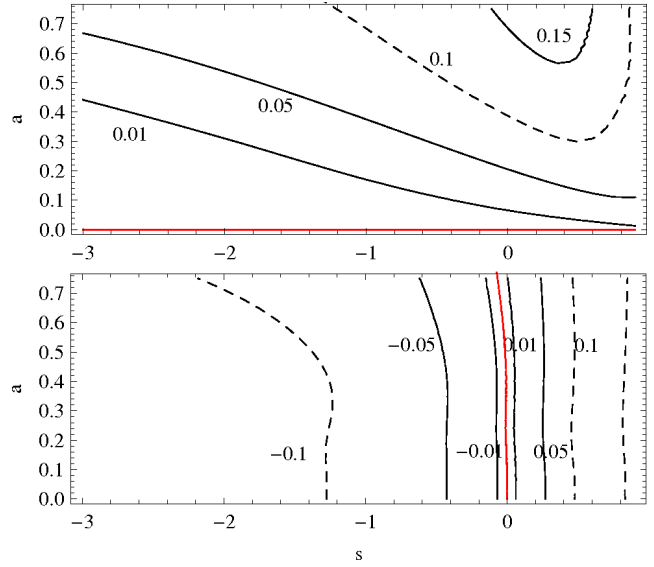}
\caption{A comparison with Gauss-Hermite moments. The upper and lower
  panels show contours of the Gauss-Hermite moments $h_3$ and $h_4$
  for our two parameter family of distributions. To guide the eye, the
  contours corresponding to $0$ and to $0.1$ are marked, respectively,
  in red and dashed lines.}
\label{GHmoments}
\end{figure}

\subsubsection{Asymmetric deviations}\label{asymm}

Asymmetric distributions can be derived from our symmetric
distributions $\LL(\s; v)$ by the transformation
\begin{equation}
\LL'(\s,\a; v)=\LL(\s; \X(\s, \a; v)).
\label{fasymm}
\end{equation} 
The asymmetric deviations are introduced by the map $v \rightarrow
\X(\s, \a; v)$.  The basic ingredients of the function $\X$ enabling
it to deliver well behaved distributions within the entire parameter
space $(\s, \a)$ are described in Appendix~\ref{asymmA}.  Here, we
only report our choice:
\begin{equation}
\X\equiv{\a\over{|\a|}}\left\{v\!-\!|\a|\ e^{\left[-\left({{v-|\a|}\over{1/2+3|\a|}}\right)^4\right]}\left({v\over{1\!+\!\s/6\!+\!|\a|}}\!+\!{{|\a|}\over2}\!-\!{3\over2}\right)\right\}\ .
\label{Xdef}
\end{equation}
While introducing asymmetries, the application of the
transformation~(\ref{Xdef}) to the symmetric distributions $\LL$ also
alters the normalization, so that in general $\LL'$ is no longer a
zero-mean, unit-integral and unit-dispersion distribution.  However,
it is straightforward to account for these matters and we define our
final two-parameters family as
\begin{equation}
\LL(\s,\a; v)\equiv{{\sigma'}\over {I'}} \LL'(\s,\a;\sigma'\cdot
v+\mu')\ ,
\label{normalization}
\end{equation}
where $\mu'$, $\sigma'$ and $I'$ are respectively the mean, dispersion
and integral of the distribution $\LL'$ in
eqn~(\ref{fasymm}). Fig~\ref{asymmexamples} displays a few examples of
asymmetric velocity distributions contained in our two-parameters
family. The different panels illustrate the asymmetric deviations
caused by positive values of $\a$ for three different values of
$\s\in\{0.5, 0, -1\}$.

\subsubsection{The Gauss-Hermite moments}\label{GHfamily}

To establish a quantitative comparison with the standard Gauss-Hermite
expansions, we measure the first two nonzero moments $h_3$ and $h_4$
of our family of distributions.  Each of them is a function of our two
parameters $\s$ and $\a$, namely
\begin{equation}
h_3  =  \a/|\a|\ H_3(\s, |\a|)\ ;\qquad\qquad
h_4  =  H_4(\s, |\a|)\ .\label{transfh4}
\end{equation}
Contours of $H_3$ and $H_4$ in the $(\s, \a)$ plane are displayed in
the upper and lower panel of Fig~\ref{GHmoments}.  In both cases, the
reference values of $0$ and $0.1$ are highlighted respectively by a
red full line and a dashed line to guide the eye.

There are some aspects worth noting.  The amount of asymmetric deviation
as quantified by $h_3$ is a function of both $\a$ and $\s$.  As is
evident from the contours of $H_3$, it is not possible to define a
one-to-one correspondence $\a\leftrightarrow h_3$. However, at least
in the vicinity of the Gaussian profile, this is almost the case for
symmetric deviations. Here, $H_4$ displays vertical contours,
characteristic of a one-to-one correspondence $s\leftrightarrow h_4$.
Nonetheless, some deviations are apparent for large negative values of
$s$.  Notice too that there are two distinct countours $H_4=0.1$,
intersecting the $\a=0$ axis at different positive values of
$\s$. Rather than being an issue for our family of distributions, this
feature is due to the inability of the Gauss-Hermite moment $h_4$ to
describe large deviations from Gaussianity. Higher Gauss-Hermite
moments are required to describe these distributions.

%
%Last point is in fact connected with a more general property. While
%still corresponding to a two-parameters family, our family of
%probability distributions contains more information than a
%Gauss-Hermite expansion truncated at $h_4$, hence with the same number
%of {\it shape parameters}. If a certain sample is best described by an
%$(\s, \a)$ pair and this provides a good statistical fit (see
%Sect.~\ref{flexcheck}), then a higher number of Gauss-Hermite
%coefficients $h_i$ must be used to attain a description similar in
%quality to the one provided by $\LL(s, a; v)$.  This highlights that a
%higher efficiency is achieved if dynamical models and observed
%datasets are compared in terms of our set of parameters.
%

%
\begin{figure}
\centering
\includegraphics[width=\columnwidth]{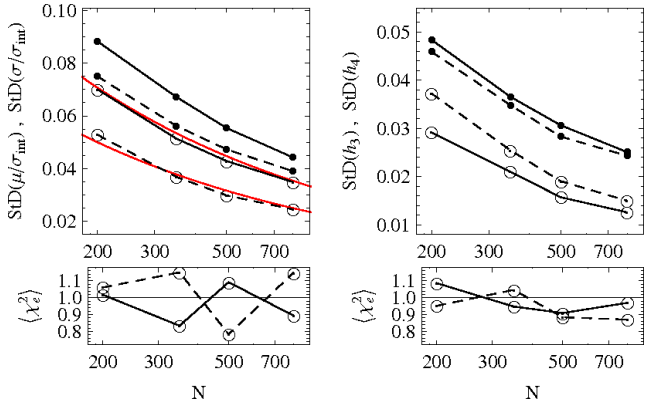}
\caption{Upper panels: Accuracy limits from limited sampling with the
  maximum likelihood method (empty circles). The standard deviation of estimated (normalized) mean $\mu$
  (full line) and dispersion $\sigma$ (dashed line) on the left; the standard deviation of
  the Gauss-Hermite moments $h_3$ (full line) and $h_4$ (dashed line) on the right. Lower panel: the
  precision test on the measurements obtained through the maximum likelihood method; coding as in the upper panels.}
\label{nnstarscomp}
\end{figure}
%
%%%%%%%%

\subsection{Tests of Accuracy and Precision}\label{tests}

To evaluate the performance of the maximum likelihood approach, we
test its accuracy and precision, in a similar manner to
Sect.~\ref{Nnoise} for the Gauss-Hermite series.  In order to
establish a direct comparison, we use as an intrinsic distribution of
the synthetic datasets a perfect Gaussian $\LL(v)=\G(\mu_{\rm int},
\sigma_{\rm int})$.  Also, we convert $\s$ and $\a$ into measurements
of $h_3$ and $h_4$ by using the transformations~(\ref{transfh4}). All
measurements are obtained by using a Metropolis-Hastings procedure,
which allows us to scan efficiently the 4-dimensional parameter space
defined by our parametrization.

Our results are collected in Fig.~\ref{nnstarscomp}.  The upper panels
display the accuracy test, and are analogous to the panels of
Fig.~\ref{nnstarsvdm}.  The results obtained for the maximum
likelihood method are denoted by empty circles.  It is evident that
both StD$(\mu)$ and StD$(\sigma)$ follow very closely their respective
statistical prescriptions ($1/\sqrt{N}$ and $1/\sqrt{2N}$, in red).
Hence, the method achieves the maximum measuring power allowed by the
sample size.  As for the deviations from Gaussianity, both StD$(h_3)$
and StD$(h_4)$ are substantially smaller than in the binned case of
Sect.~\ref{Nnoise}. Experiments with non-Gaussian intrinsic velocity distributions
show an even smaller shot noise, athough with qualitatively similar figures. 
The relative gain in accuracy for the detections
of symmetric deviations is found to be an increasing function of the
sample size, reaching approximately 2 for $N=800$ and surpassing 2 for
asymmetric deviations.  These quantities refer to the idealized case
of datasets with no observational uncertainty ($\delta_i=0$) and
uniform certainty of membership ($p_i=1$). Hence, they represent only
lower bounds for the actual gains that are achievable in any real
case.

Lower panels display the precision test, which evaluates the
reliability of the uncertainties returned by the maximum likelihood
procedure.  The $\langle \chi^2_e\rangle$ quantity in the plots
represents the average (over the number of performed tests) for the
quantity
\begin{equation}
\chi^2_e=\left({{\theta-\theta_{\rm int}}\over{e_{\theta}}}\right)^2\ ,
\label{chiplot}
\end{equation}
where $\theta$ stands for any parameter of the family of distributions
and $\theta_{\rm int}$ is its intrinsic, input value.  We recall that
$e_{\theta}$ denotes the uncertainty on the measured value of the
parameter as returned by the marginalized likelihood\footnote{These
  uncertainties are not symmetric in general, so eqn~(\ref{chiplot})
  is calculated by using the relevant higher or lower limit of the
  68\% confidence interval, depending on whether the best fitting
  value is larger or smaller than the intrinsic one.}.  We find that
the errors on $\mu$ and $\sigma$, as well as those on $s$ and $a$
behave as desired, with all $\langle \chi^2_e\rangle$ averaging
approximately to the expected value of unity.

\begin{figure}
\centering
\includegraphics[width=.9\columnwidth]{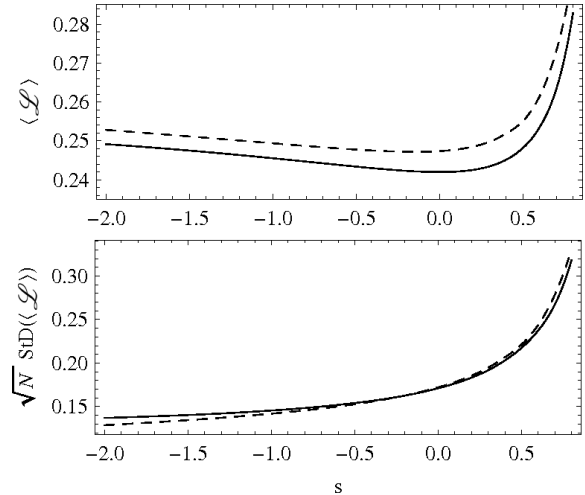}
\caption{The average $\langle\LL\rangle$ (upper panel) and standard
  deviation StD$(\langle\LL\rangle)$ (lower panel) for different
  symmetric deviations $s$. In both panels, full lines represent the
  case $(s, a=0)$, while dashed lines are $(s, |a|=0.5)$.}
\label{statquant}
\end{figure}

%%%%%%%%
\subsection{A Check on the Degree of Flexibility}\label{flexcheck}

It is natural to raise the question: {\it what if the intrinsic
  distribution is not included in our two-parameter family?} This may
represent the greatest disadvantage of the maximum likelihood
implementation, because for extremely high sample sizes and
observational precision of kinematic measurements, the standard
Gauss-Hermite expansion can be made as flexible as necessary by adding
higher order terms. However, it is possible to set up an efficient
device that controls whether the family of distributions is
appropriate and flexible enough.

For the Gauss-Hermite series, this device is represented by
Myller-Lebedeff's theorem, which equates the integral of the residuals
between the observed distribution and its best Gaussian fit with the
sum of the Gauss-Hermite moments themselves (eqns~(12-14)
in~\citet{vdMF93}). This allows us to check whether the adopted
truncation of the Gauss-Hermite series is completely satisfactory, and
if further higher order terms are required.

Within our maximum likelihood approach, it is necessary to ask a more
purely statistical question: given the observed sample $\vec{V}$
(which comes together with the associated uncertainties $\vec\Delta$,
the membership probabilities $\vec{P}$, and its sample size $N$) and
the distribution $\LL$ that -- within the considered family --
provides the maximum likelihood, {\it would an analogous sample,
  actually extracted from the same $\LL$, be fitted significantly
  better?}  It is possible to answer this question quite easily in an
analytic way.

\begin{figure}
\centering
\includegraphics[width=.8\columnwidth]{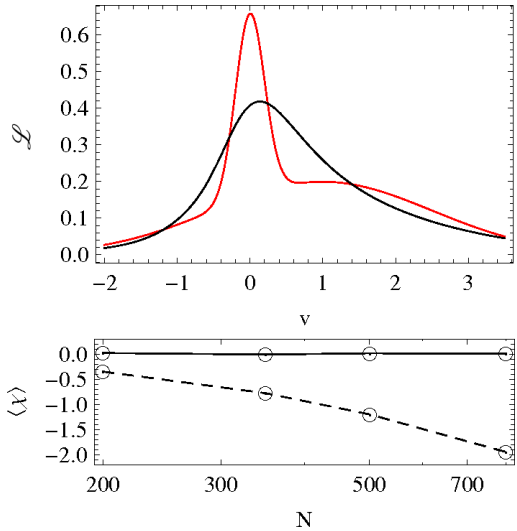}
\caption{A test of our statistical device. Upper panel: the input
  distribution in red and its best description within our family in
  black. Lower panel: evolution with the sample size of the average
  (on the series of tests) of the $\chi$ quantity defined in
  eqn~(\ref{chilik}). Dashed line: the synthetic datasets are distributed  
  according to the red input distribution shown in the upper panel; full line: the synthetic datasets are distributed  
  according to the black best-fit distribution shown in the upper panel.}
\label{chitest}
\end{figure}

Let us suppose that $\vec\Theta$ is the set of parameters that
provides the maximum likelihood for the sample $\vec{V}$, accompanied
by $\vec\Delta$ and $\vec{P}$:
\begin{equation}
L(\vec\Theta)= \prod_{i=1}^{N}p_i\ \left[\LL \ast \G(0, \delta_i)\right](v_i)\ .
\label{mlik}
\end{equation}
The value of $L(\vec\Theta)$ can be compared with the characteristic
value of the analogous product in eqn~(\ref{mlik}) in which the
velocities $v_i$ {\it are actually extracted from the distribution
  given by the set of $\vec\Theta$}:
\begin{equation}
\langle L(\vec\Theta)\rangle = \langle\prod_{i=1}^{N}p_i\ \LL \ast \G \rangle= \prod_{i=1}^{N}p_i\ \int\left[\LL\ast\G(0,\delta_i)\right]^2\ .
\label{mliktheo}
\end{equation}
Since both quantities defined by eqns~(\ref{mlik})
and~(\ref{mliktheo}) converge to zero quickly with $N$, we find it
more convenient to consider their nonvanishing counterparts
$\sqrt[N]{L(\vec\Theta)}$ and $\langle
\sqrt[N]{L(\vec\Theta)}\rangle$.  In order to ease the notation, we
use the simplification
\begin{equation}
\left\{\begin{array}{rcl} \langle \sqrt[N]{L(\vec\Theta)}\rangle & = &
\langle L \rangle\ \\ \sqrt[N]{L(\vec\Theta)}& = & \bar{L}
\end{array}\right. \ .
\end{equation} 
If the distance between $\bar L$ and $\langle L \rangle$ can be
accounted for by the natural scatter introduced by the sample size $N$
only, then the distribution given by the set $\vec\Theta$ provides a
statistically perfect description of the sample $\vec{V}$.  This
natural scatter is clearly given by
\begin{equation}
{\rm StD}\left[\langle L \rangle\right]= \sqrt{\langle
  {L(\vec\Theta)^{2/N}}\rangle-\langle L \rangle^2}\ ,
\label{stdmlik}
\end{equation}
hence the quantity we are interested in is 
\begin{equation}
\chi={{\left(\bar L-\langle L \rangle\right)}/{{\rm StD}\left[\langle
      L \rangle\right]}}\ .
\label{chilik}
\end{equation}
Values of $\chi^2$ up to unity indicate that the sample $\vec V$ is
statistically well described.  Negative values of $\chi$, with
absolute value significantly larger than unity, indicate that the
adopted parametrization is not able to provide a good statistical
description of the sample.

To apply this criterion, we need explicit expressions for both
$\langle L\rangle$ and ${\rm StD}\left[\langle L\rangle\right]$.  It
is useful to note that for a fixed set of parameters $\vec\Theta$,
both $\langle L\rangle$ and ${\rm StD}\left[\langle L\rangle\right]$
are invariant with respect to a change in $\mu$, which we can
ignore. Also, if we indicate with $\langle L(\sigma=1,
\vec{\Theta}_{\rm sh})^{1/N}\rangle$ and ${\rm StD}\left[\langle
  L(\sigma=1, \vec{\Theta}_{\rm sh})^{1/ N}\rangle\right]$, the values
attained for $\sigma=1$ (all others $\theta_j$, $\vec\Delta$ and
$\vec{P}$ fixed), then for a general $\sigma$ it is easy to verify
that
\begin{eqnarray}
\langle L(\sigma, \vec{\Theta}_{\rm sh})^{1/N}\rangle & = & \langle L(\sigma=1, \vec{\Theta}_{\rm sh})^{1/N}\rangle\big/\sigma\\
{\rm StD}\left[\langle L(\sigma, \vec{\Theta}_{\rm sh})^{1/ N}\rangle \right]& = & {\rm StD}\left[\langle L(\sigma=1, \vec{\Theta}_{\rm sh})^{1/ N}\rangle\right] \big/\sigma\ ,\nonumber
\label{sigmascaling}
\end{eqnarray}
where the uncertainties are scaled accordingly, i.e.,
$\delta_i\rightarrow\delta_i/\sigma$.  As a consequence, we can
restrict the problem to the case $\sigma=1$.

We use now the fact that, for large $N$, the different convolutions
can be accounted for by the mean $\delta_m$ of the sample of
uncertainties $\vec\Delta$, and after some algebra, we find the
following asymptotic expressions, valid for the case $\sigma=1$:
\begin{eqnarray}
\langle L\rangle & = & \prod_i
p_{i}^{1/N}\int\left[\LL\ast\G(0,\delta_m/\sigma)\right]^{1+1/N}\nonumber\\ &
= & p_{\rm gm}\ \exp(A) + O(1/N)
\label{implint1}\\
{\rm StD}\left[\langle L\rangle\right] & = & \left\{\prod_i
p_{i}^{2/N}\int\left[\LL\ast\G(0,\delta_m/\sigma)\right]^{1+2/N} -
\langle L\rangle^2\right\}^{1/2}\nonumber\\ & = & p_{\rm
  gm}\ \exp(A)\sqrt{B-A^2}/\sqrt{N} + O(1/N)
\label{implint2}
\end{eqnarray}
in which $p_{\rm gm}$ is the geometric mean of the sample's membership
probabilities $\vec{P}$
\begin{equation}
p_{\rm gm}=\sqrt[N]{\prod_{i=1}^N p_i}
\label{geommean}
\end{equation}
and $A$ and $B$ are the following simple integrals (we implicitly
assume that $\G=\G(0,\delta_m/\sigma)$)
\begin{eqnarray}
A = A(\sigma=1,\vec{\Theta}_{\rm sh}) & = & \int (\LL\ast\G)\log\left(\LL\ast\G\right)\label{explint1}\\
B = B(\sigma=1, \vec{\Theta}_{\rm sh}) & = & \int (\LL\ast\G)\log\left(\LL\ast\G\right)^2\ .
\label{explint2}
\end{eqnarray}
Eqns.~(\ref{implint1} - \ref{explint2}) allow us to compute directly,
for any value of the parameters $\theta_j$, all necessary ingredients
to compute $\chi$ in eqn~(\ref{chilik}), and hence to understand
whether the fit of the observed sample is indeed statistically good.

As a reference and an example, we consider here the simplified case in
which there is no observational uncertainty, $\delta_m=0$, and the
likelihood is maximized by the distribution $\LL$, having dispersion
$\sigma=1$.  With a slight abuse of notation, we use
$\langle\LL\rangle$ to indicate the average of the likelihood in the
sense of eqn~(\ref{mliktheo}) and ${\rm StD}(\langle\LL\rangle)$ to
indicate its standard deviation, as in eqn~(\ref{stdmlik}).  For the
Gaussian case, $\LL=\G(\mu, \sigma=1)$, both integrals $A$ and $B$ are
analytic and we find
\begin{equation}
\begin{array}{rcccl}
\langle\G\rangle & = & 1\big/\sqrt{2\pi e}   & \approx &  0.24197\ ;\\
\sqrt{N}\ {\rm StD}(\langle\G\rangle)    & = & 1\big/\sqrt{4\pi e}  & \approx &  0.1711\ .
\end{array} 
\label{gaussexact}
\end{equation}
Deviations in $\LL$ from the Gaussian profile determine deviations in
the average $\langle\LL\rangle$ as well as in the corresponding
standard deviation ${\rm
  StD}(\langle\LL\rangle)$. Fig.~\ref{statquant} displays the
behaviour of $\langle\LL\rangle$ and $\sqrt{N}{\rm
  StD}(\langle\LL\rangle)$ for the cases $(\s, \a=0)$ (full line) and
$(\s, |\a|=0.5)$ (dashed line). Both the displayed quantities increase
significantly for positive values of $s$, due to the change in shape
of the associated distributions.

Finally, Figure~\ref{chitest} illustrates a practical test. The
distribution in red in the upper panel is used to produce synthetic
datasets of different sample sizes, which are then fed to our maximum
likelihood formalism. This distribution is not included in our
parametric family, and, as a comparison, the distribution in black in
the same panel displays its best fit. We perform a large number of
tests for different sample sizes and record the evolution of the
average of the quantity $\chi$, computed at each test, in the lower
panel of the same Figure, as a dashed line. For small sample sizes, it
is almost impossible to distinguish the two
distributions. Nonetheless, as the sample size increases, the
properties of the input distribution become more evident and cannot be
completely reproduced within our family, so that $\chi$ reaches a
value of $-2$ for $N=800$. The black full line in the lower panel
shows, for comparison, the average of the $\chi$ quantity that we
obtain for synthetic samples that are drawn directly from the best fit
distribution, and that average to the expected value of zero for any
sample size.
 
\begin{figure}
\centering
\includegraphics[width=0.8\columnwidth]{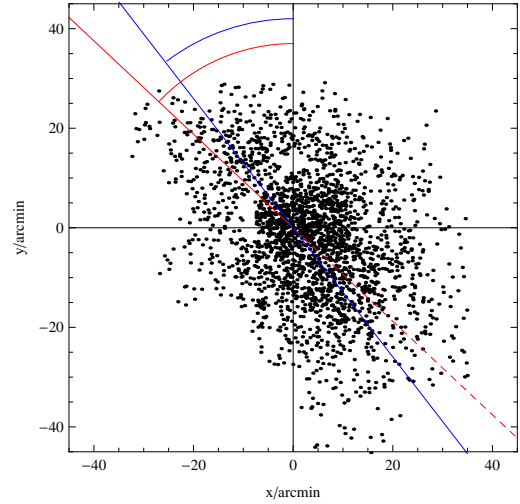}
\caption{The positional distribution of kinematic measurements in
  Fornax as from the dataset presented by \citet{Wa09data}. The position
  angle $\theta_{\rm MA}$ is displayed in red, while the apparent
  rotation axis $\theta_{\rm app}$ \citep{Pi07} is shown in blue.}
\label{fornax2d}
\end{figure}
\begin{figure*}
\centering
\includegraphics[width=\textwidth]{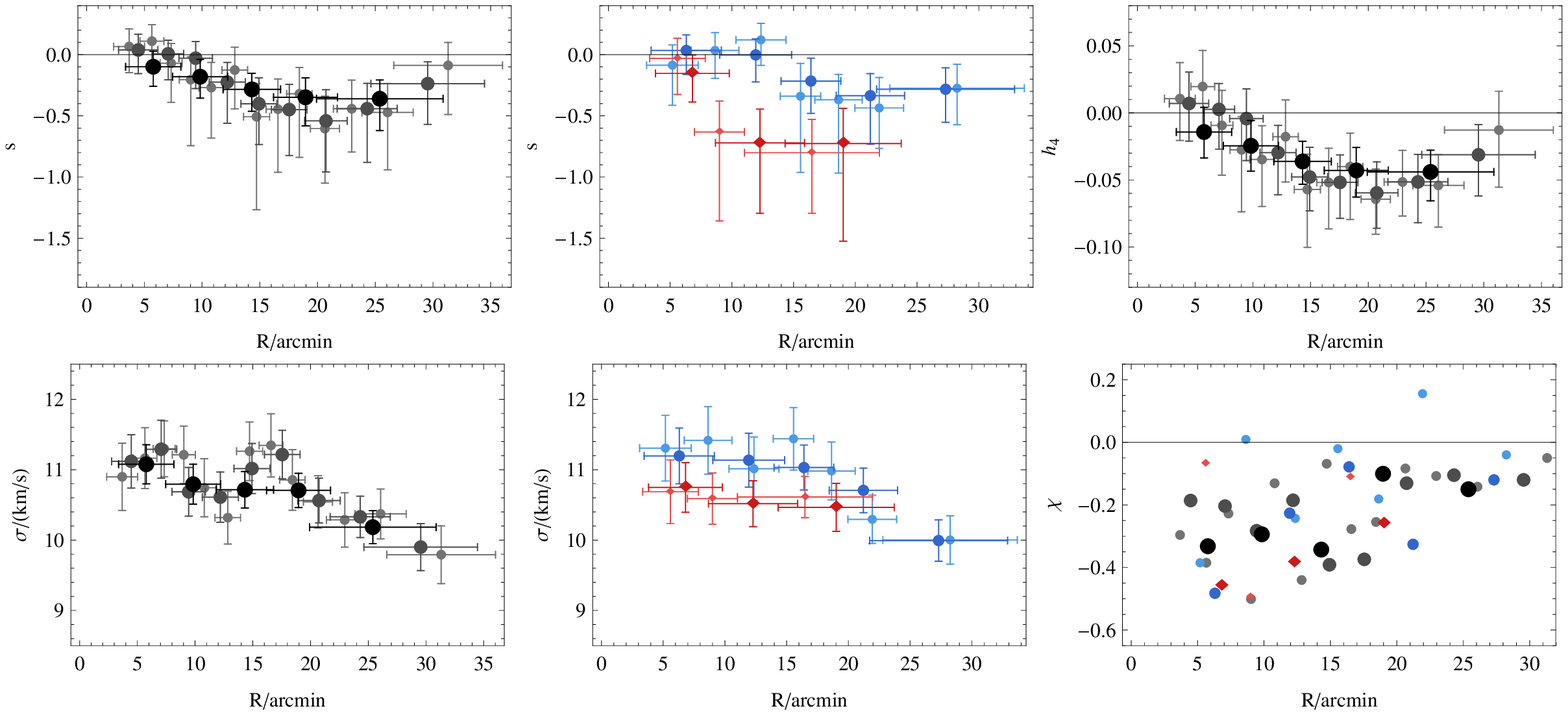}
\caption{Results for the Fornax dSph: circular annuli.  First row:
  the symmetric deviations.  Second row, left and middle panels: the
  dispersion.  Second row, right panel: the quantity $\chi$ for all
  displayed maximum likelihood measurements.}
\label{fornaxsym}
\end{figure*}
\begin{figure}
\centering
\includegraphics[width=.9\columnwidth]{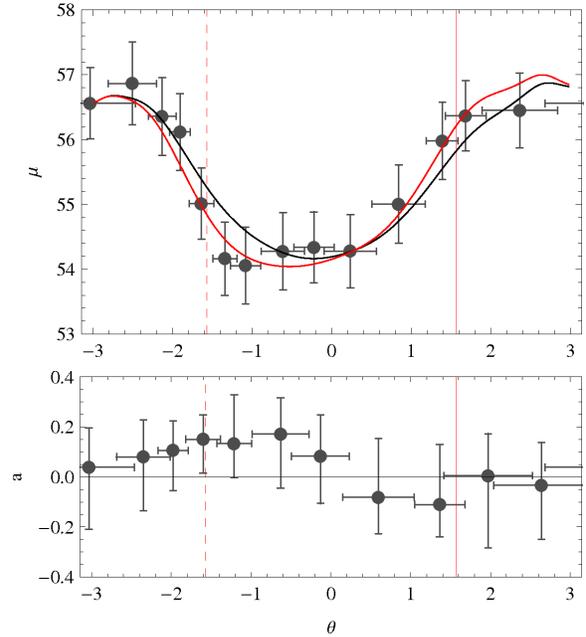}
\caption{Results for the Fornax dSph: angular sectors.  Upper panel:
  datapoints display the mean $\mu$; the black full line shows the
  prediction for $\mu$ determined by the astrometric measurement of
  proper motion measured \citep{Pi07} under the assumption of no
  streaming motion; the red full line displays the correction
  introduced by intrinsic rotation. Lower panel: the asymmetric
  deviations.  }
\label{fornaxang}
\end{figure}
%

%%%%%%%%%%%%%%%%%%%%%
\section{Applications: The Galactic dSphs}\label{appl}

\subsection{Fornax}

The kinematic sample presented by~\citet{Wa09data} consists of 2409
measurements for stars with a probability of membership higher than
0.9. We include this information in our likelihood~(\ref{lik}), and we
discard measurements with a smaller membership probability. As already
found in Section~\ref{attenuation}, this kinematic sample comes with a
(normalized) level of uncertainty of $\delta_m/\sigma\approx 0.22$,
which is the smallest in the currently available selection of dSphs.
To construct our set of observables, we transform the coordinates of
the stars in the plane of the sky with
\begin{eqnarray}
x_i & = & - c \ \cos\delta_i\ \sin(\alpha_i-\alpha_0)\\ y_i & = &
c\ \left[\sin\delta_i \cos\delta_0 - \cos\delta_i \sin\delta_0
  \cos(\alpha_i-\alpha_0)\right],
\label{cartesiantransf}
\end{eqnarray}
in which $c$ is a constant ($=10800/\pi$ for coordinates in
arcminutes). We adopt the coordinates (J2000) of Fornax's center as
in~\citet{Ma98}, $(\alpha_0, \delta_0)=(2^h 39^m 59^s, -34^\circ
27.0')$. The photometric ellipticity ($e=0.30$) and the major axis
position angle ($\theta_{\rm MA}=46.8^\circ$) are taken
from~\citep{Ba06}.  Fig.~\ref{fornax2d} shows the resulting spatial
distribution of the kinematic measurements on the plane of the sky,
with the angle $\theta_{\rm MA}$ highlighted in red.

\subsubsection{Symmetric deviations}

Given the number of available kinematic measurements and the results
of our accuracy tests, we consider a set of different subsample sizes
for our measurements: $N\in\{350, 500, 800\}$. We experimented with
both circular and elliptical annuli, but we have found no significant
difference, and hence report results for the circular annuli only. For
a comparison, we also consider results for minor axis and a major axis
regions. Each of these is defined as the sum of the two opposite
Cartesian quadrants centered on the relevant axis. Given the smaller
number of stars in each of these regions, only $N\in\{350, 500\}$ were
considered. As we demonstrated in Sect.~2.3, our measurements of the
line profiles in circular annuli are not affected by apparent
rotation.

Results for symmetric deviations $s$ and velocity dispersion $\sigma$
are displayed in the upper and lower panels of Fig~\ref{fornaxsym}.
In both rows, the left panels (in shades of grey) illustrate the
results for circular annuli, while middle panels show the division
according to the major and minor axes regions (respectively, in
shades of blue and red). Different shades of the same colour, and
different sizes of the corresponding datapoint, are used to indicate
the sample size used for each measurement, with larger sizes
associated to darker and larger datapoints. The upper-right panel
translates the symmetric deviations in terms of the Gauss Hermite
coefficient $h_4$. The lower-right panel displays the value of the
quantity $\chi$ corresponding to each single maximum likelihood
measurement.

The symmetric deviations display a clear evolution from positive
values of $s$ (and $h_4$) in the center of Fornax, to negative values
of $s$ (and $h_4$) at larger radii, with a tentative return towards a
more Gaussian profile at the end of the sample`s radial coverage
($\approx 3$ half-light radii).  The major and minor axes regions show
some differences and the minor axis region displays a stronger
(negative) signal for $s$.  This may perhaps be consistent with an
elliptical kinematic pattern following the isophotes.  Nonetheless it
is interesting to notice that most of the signal for flat-topped
distributions is indeed coming from the minor axis region.  We confirm
a mildly declining dispersion profile in Fornax and also note a
systematic difference between the major axis and minor axis regions,
with the minor axis showing a lower line of sight velocity dispersion.

If we were to interpret the result by following the suggestions of
both~\citet{Ge93} and~\citet{vdMF93}, we would conclude that Fornax
shows some degree of tangential anisotropy, at least outside its
central regions.  We notice that the tentative `peak' of positive
values of $s$ (and $h_4$) in the very centre may be interpreted in
different, possibly non exclusive, ways. The first interpretation
invokes the `complementarity property', recognized by \citet{De87}: a
tangentially biased structure has flat-topped ($s<0$) distributions
outside some transition radius and a more spiky ($s<0$) distribution
in the central regions. The second interpretation is the existence of
two populations with distinct kinematic properties. It is easy to see
that the superposition of two approximate Gaussians with different
widths would be recognized as a distribution with a positive $s$ --
the exact value of which is dependent on both the ratio of numbers and
dispersions of the two superposing populations. Finally, there may be
place to accommodate a central intermediate mass black hole (IMBH). In
this respect, it would be interesting to compare detailed modelling of
our results with the constraints obtained by \citet{Ja12} for the
IMBH's mass.  Regarding the issue of multiple stellar populations, we
also note a systematic tendency of measurements in the inner parts of
the system to have higher values of $\chi$. This is consistent with
the fact that while at larger radii we are effectively modelling just
the metal poor stellar population, towards the center we register the
effects coming from two superposing populations.

We do not report explicit results for the asymmetric deviations $a$ or
for the mean $\mu$ for circular annuli. Unlike symmetric deviations,
asymmetric deviations average to zero over circular annuli. Our
results confirm this expectation and we prefer to address the
characterization of asymmetric deviations by considering a purely
angular subdivision of the dataset, using angular sectors that make no
reference to the distance from the centre.

\subsubsection{Asymmetric deviations and apparent rotation}

In the Fornax dSph, the astrometrically derived proper motion
\citep{Pi07} agrees at approximately 1-sigma with the proper motion
deduced using the kinematic data under the assumption of no streaming
motions \citep{Wa08pm}. This testifies to the fact that, if any
intrinsic rotation is present, it must be small by comparison with the
velocity field given by the apparent rotation.  However, a precise
measurement for both proper motion and rotation in dSphs is relevant
for comparison with simulations, and for constraining the formation
history of such systems.  For this reason, we reconsider this issue
here, and note that our ability to measure asymmetries in the line of
sight profiles can help us constrain the intrinsic rotation
field. This is because, in the plane of the sky, asymmetries and
intrinsic rotational velocities are likely to be strongly correlated.

In the lower panel of Fig.~\ref{fornaxang}, we measure the the
"asymmetry field" in angular sectors $a(\theta)$ around Fornax's
center (on subsamples $N=400$). Although almost everyhere nearly zero,
we do detect a $2\pi$-periodic signal, which is indeed compatible with
an intrinsic rotation.  Not all datapoints in the panel are
independent, and hence we do not try to fit our result, but it is
encouraging that the peaks of the signal are approximately aligned
with the major axis of the system, which is displayed as red vertical
lines, $\theta_{\rm peak}\approx \theta_{\rm MA}$. This signal is not
due to apparent rotation, and is robust against subtraction of the
apparent rotation field.

The datapoints in the upper panel of Fig.~\ref{fornaxang} show the
associated mean in angular sectors $\mu(\theta)$ (on subsamples
N=350). This can be compared with the black full line in the same
panel, which displays the apparent rotation that the astrometrically
derived proper motion implies on the 2-dimensional distribution of
kinematic measurements.  It is clear that any disagreement is again
correlated with the position of the major axis, and has opposite signs
in opposite directions, in a way that is compatible with intrinsic
rotation. Unfortunately, despite the quality of the dataset, it is not possible
to derive a statistically meaningful characterization of the two
dimensional intrinsic velocity field. This implies, approximately, an intrinsic rotation of
about 1 kms$^{-1}$ for the outermost tracers aligned with the major
axis in either directions.

\begin{figure}
\centering
\includegraphics[width=\columnwidth]{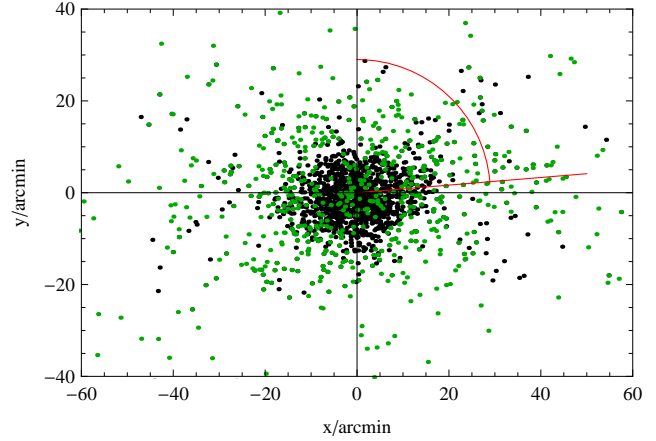}
\caption{The distribution of kinematic measurements in Sculptor as
  from the dataset presented by \citet{Wa09data} (in black) and from the
  dataset presented by \citet{St10} in green. The position angle
  $\theta_{\rm MA}$ is displayed in red.}
\label{sculptor2d}
\end{figure}
\begin{figure*}
\centering
\includegraphics[width=\textwidth]{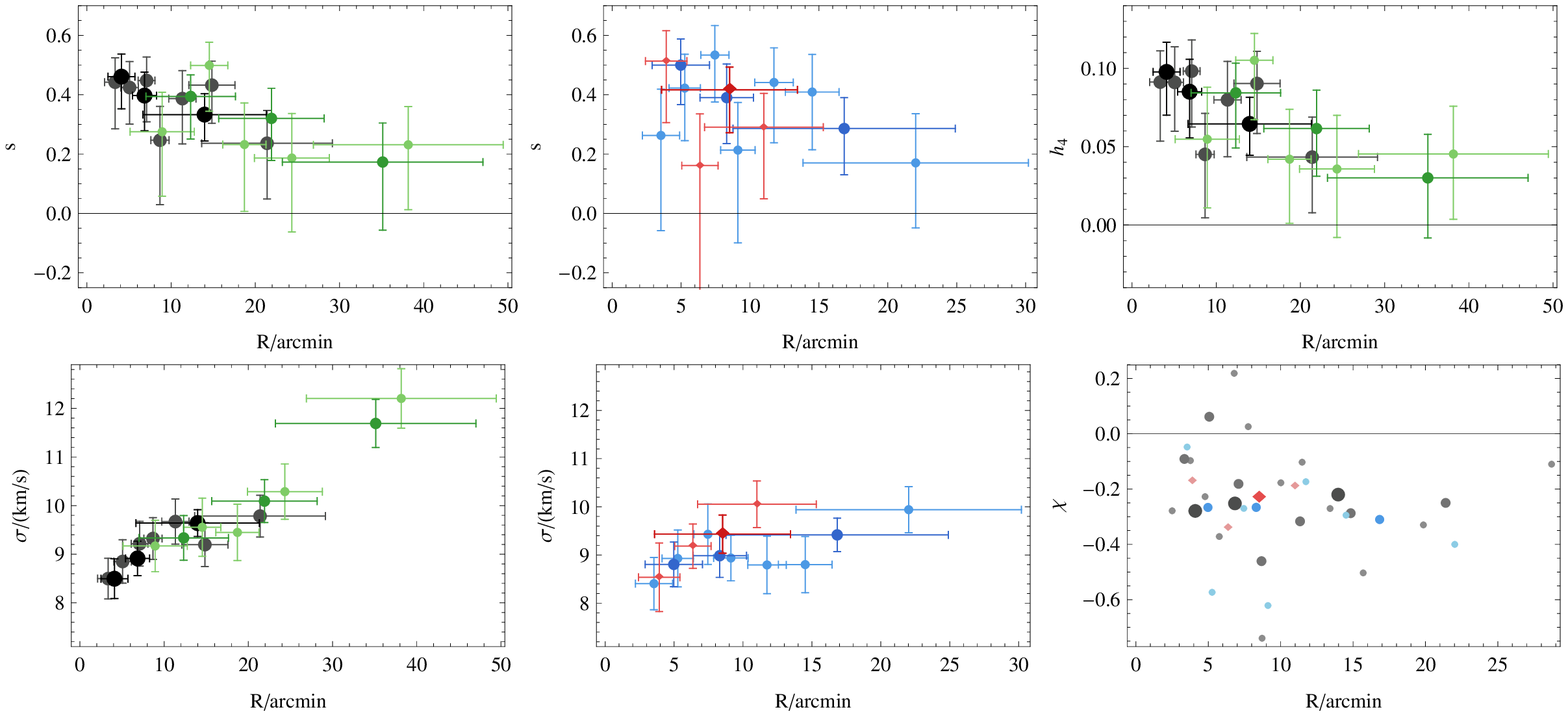}
\caption{Results for the Sculptor dSph: circular annuli.  First row:
  the symmetric deviations.  Second row, left and middle panels: the
  dispersion.  Second row, right panel: the quantity $\chi$ for all
  displayed maximum likelihood measurements.}
\label{sculptorsym}
\end{figure*}
\begin{figure}
\centering
\includegraphics[width=.9\columnwidth]{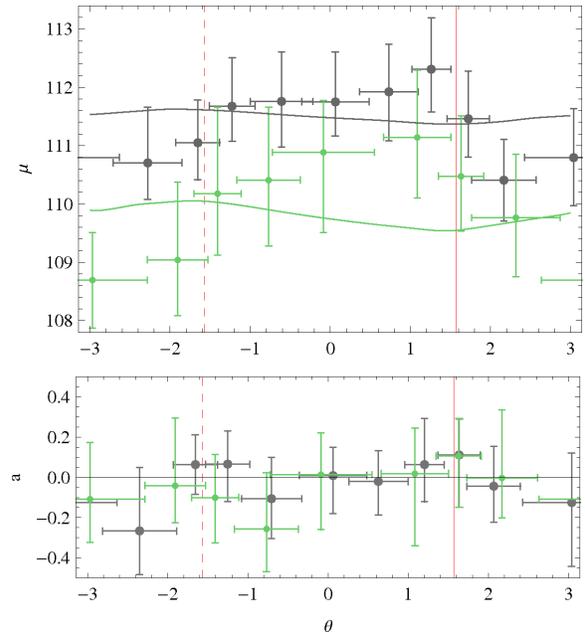}
\caption{Results for the Sculptor dSph: angular sectors.  Upper
  panel: datapoints display the mean $\mu$; full lines show the
  prediction for the mean $\mu$ determined by the astrometric
  measurement of proper motion measured \citep{Pi06} and no intrinsic
  rotation (systematic velocities as from the datasamples).  Lower
  panel: the asymmetric deviations.  }
\label{sculptorang}
\end{figure}

\subsection{Sculptor}

The available sample from \citet{Wa09data} contains 1370 line of sight
velocity measurements with membership probability higher than 0.5. We
adopt the coordinates (J2000) of the dSph`s centre from
\citet{Ma98}: $(\alpha_0, \delta_0)=(1^h 00^m 09^s, -33^\circ 42.5')$.
The photometric ellipticity $e=0.26\pm 0.01$ and position angle
$\theta_{\rm MA}=-85.3^\circ\pm 0.9^\circ$ are taken
from~\citet{deB11}.  As reported in Section~\ref{attenuation}, this
kinematic sample comes with a (normalized) level of uncertainty of
$\delta_m/\sigma\approx 0.325$.  We accompany this kinematic sample
with the one provided by \citet{Ba08} and then recalibrated by
\citet{St10} (results related with this datasets are displayed in green in all relevant Figures). 
To avoid misalignment between the catalogs,
for this second dataset we use the dSph center as determined
in~\citet{deB11}. Even though the number of line of sight velocity
measurements is lower with 629 giants, they cover a significantly more
extended radial region (see Fig~\ref{sculptor2d}), which makes the two
datasets complementary. Also, a smaller (normalized) level of
uncertainty $\delta_m/\sigma\approx 0.17$ is achieved.

Given the reduced number of kinematic tracers in comparison to Fornax,
we are forced to consider smaller sample sizes: $N\in\{350,
500\}$ for the circular annuli and $N\in\{200,350\}$ for the major
axis and minor axis regions. Results for symmetric deviations $s$ and
velocity dispersion $\sigma$ are displayed respectively in the upper
and lower panels of Fig~\ref{sculptorsym}. The collected results refer
to circular annuli, and again no significant differences were found
for the case of elliptical annuli.

The symmetric deviations display a marked preference for positive
values of $s$ (and $h_4$) for the entire radial range covered by the
tracers. This behaviour is confirmed by both datasets, which are found
in perfect agreement.  Identical profiles (within the uncertainties)
are found for the major axis and minor axis regions. Taken at face
value, these results support a radially biased dynamical structure,
which was also the result of~\citet{Am12}. The central peak in $s$ as
well as the tendency for higher values of $\chi$ towards the center
mimic the case of Fornax, and hence point towards the effect of
superposing populations, even though we cannot exclude other dynamical
origins.

We confirm the outwardly increasing velocity dispersion profile in
Sculptor, although it should be remembered that the dataset from
\citet{Ba08} is not provided with probabilities of membership, hence
the outermost points may probably be affected by
contamination. Nonetheless, in the radial range where both datasets
are available, they agree very well.  Surprisingly, we find that the
two datasets do not agree in the deduced means $\mu$. The upper panel
of Fig.~\ref{sculptorang} displays the mean in angular sectors
$\mu(\theta)$. We detect very similar angular behaviour, but the two
datasets seem to be shifted uniformly of about 1 kms${}^{-1}$. Given
that higher order moments all agree, we suspect that such a significant
difference may be systematic in origin. Therefore, particular caution 
should be used when attempting to merge the two datasets.

Unfortunately, neither of the two datasets displays a conclusive
signal for the asymmetric deviations (see lower panel in
Fig.~\ref{sculptorang}), which does not allow us to make progress in
the determination of Sculptor`s proper motion or intrinsic rotation.
It is known that the astrometric measurement of Sculptor`s proper
motion \citep{Pi06} does not agree with the kinematic rotation
signal. This suggest either the presence of significant intrinsic
rotation or perhaps an error in the astrometric measurement, and is
exemplified by the disagreement between the datapoints and the full
curves in the upper panel of Fig.~\ref{sculptorang}. Such curves,
display the apparent rotation implied by Piatek's proper motion, and
have been normalized to the systematic velocity derived separately by
each dataset. In turn, even though roughly agreeing on its direction and qualitatively with 
the rotation identified in \citet{Ba08}, the two kinematic samples would suggest proper motions of different
magnitudes, hence further observational effort will be necessary for
progress. 

\begin{figure*}
\centering
\includegraphics[width=\textwidth]{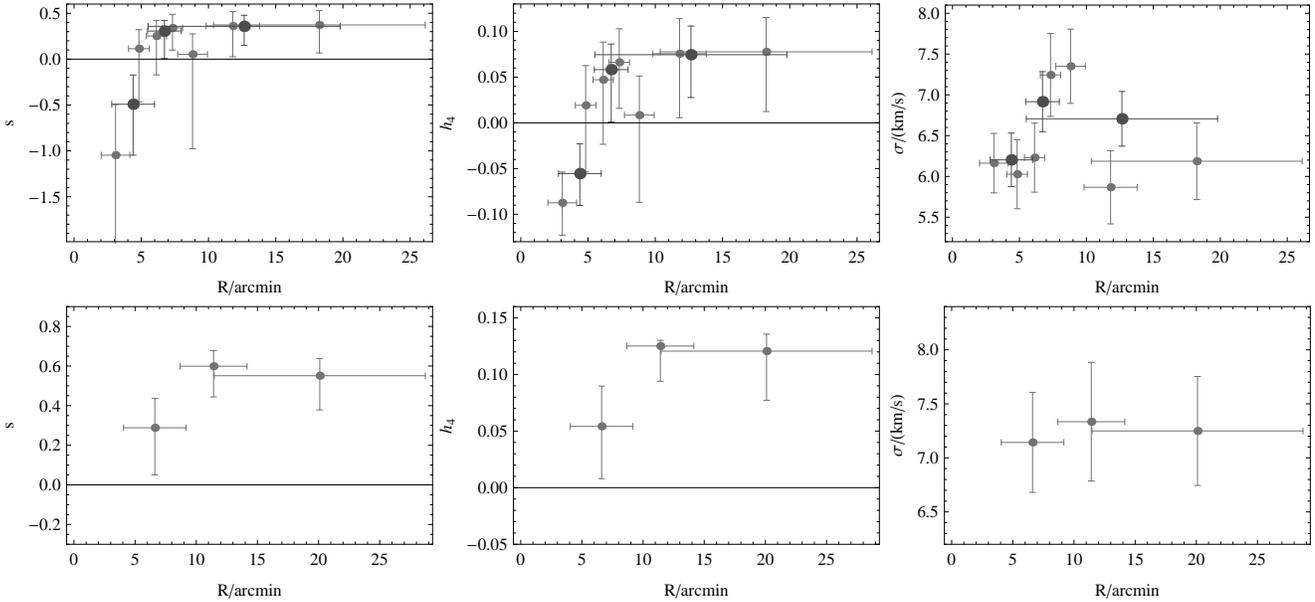}
\caption{Results for the Carina (upper panels) and the Sextans (lower
panels) dSphs. Left and middle panels in both rows: results for the
asymmetric deviations.  Right panels: velocity dispersion.}
\label{carsext}
\end{figure*}
\subsection{Carina and Sextans}

The sizes of the kinematic samples regarding the Sextans and Carina
dSphs are significantly smaller than the two previous cases,
respectively with 449 and 780 stars with membership probability higher
than 0.5 (424 and 758 higher than 0.9). Also, since the intrinsic line
of sight velocity dispersion is smaller than in Fornax and Sculptor,
the average level of uncertainty goes up, as reported in
eqns~(\ref{errlevels}). Since the apparent velocity fields are poorly
constrained, we decide to avoid their subtraction. For completeness,
we use the coordinates of the centers of these dSphs as listed in
\citet{Ma98}. We use, respectively for Sextans and Carina, sample
sizes of $N=200$ and $N\in\{175, 325\}$.

Results for the symmetric deviations in circular annuli for both dSphs
are shown in Fig.~\ref{carsext}. Both systems show a tendency (outside
the innermost regions in the case of Carina) for line of sight
distributions that are more peaked than Gaussian, which is compatible
with a radial bias in their orbital structure.

%%%%%%%%%%%%%%%%%%%%%

\section{Conclusions}\label{concl}

We have devised an efficient method to extract the shape information
for line profiles of discrete kinematic data. Such information is
often crucial in constraining the orbital structure of stellar
systems.  Independent knowledge of the orbital anisotropy is necessary
to break the mass-anisotropy degeneracy, and hence to constrain the
mass density profile. Clear-cut determination of the mass profile at
both very small and large radii is made challenging by such
degeneracies, but nonetheless provides a crucial test for our picture
of galaxy formation. Also, the orbital structure retains memory of the
initial conditions in which the tracers were formed, and so constrains
albeit indirectly the different galaxy formation mechanisms.

Our methods are complementary to the standard Gauss-Hermite formalism,
that is best suited for continuous data obtained from absorption line
spectra. Discrete kinematic measurements are affected by inhomogeneous
uncertainties and often come with varied probabilities of
membership. The Gauss-Hermite formalism is unable to account for all
this different information, in contrast to a Bayesian approach, which
also allows us to avoid any binning procedure.

Since the Gauss-Hermite series is not positive definite, it cannot be
used as a probability distributions. Instead, we construct a new
family of line profiles derived from velocity distributions and use
them in the context of Bayesian inference. Our family has two
parameters, namely $s$ which quantifies symmetric deviations from the
Gaussian profile and $a$ which refers to the asymmetric deviations.
The parameter $s$ has a kinematic interpretation, and is associated
with line profiles that mimic those of constant anisotropy models
(with an exponential dependence on the energy).

Our methods allow us to measure directly the intrinsic line profiles
$\LL$, rather than the profile convolved with the observational
uncertainties.  The advantage of such an approach is substantial. Any
signal of a deviation from Gaussianity is significantly stronger in
the intrinsic line profile. Hence, a smaller sample size is sufficient
to reach the level at which signal itself is larger than the shot
noise. We quantify the magnitude of this noise as a function of the
sample size and find that, within the Gauss-Hermite formalism, this
noise is higher than the expected signal in both $h_3$ and $h_4$ for
sample sizes smaller than N$\approx$200.  This casts doubts on measurements
obtained on significantly smaller sample sizes, especially if important observational
uncertainties are present. We find that our maximum likelihood methods
perform in comparison systematically better. Even in the case of zero
uncertainties, we achieve a relative gain in accuracy that is about 2
on $h_4$ (for sample sizes $N=800$) and higher for $h_3$. These
quantities cannot but improve in presence of observational
uncertainties and estimates for the probabilities of membership.

To ensure that our methods can give reliable descriptions of the
shape, we present a simple test that is able to assess the statistical
quality of the fit. This is obtained by quantifying the scatter that
limited sampling implies on the average value of the likelihood, which
can be done analytically. We apply this test to a practical example
and confirm that it is indeed able to identify cases in which the
adopted family of line profile is not able to provide a good
statistical representation of the data.

We apply our formalism to the discrete velocity datasets of the dwarf
spheroidals of the Milky Way. We quantify the effects of apparent
rotation due to systematic proper motions and find that these are not
an issue for the dSphs.  We measure detailed radial profiles for the
symmetric deviations in Fornax, Sculptor, Carina and Sextans. All
systems but Fornax are characterized by line of sight profiles that
are substantially more peaked than Gaussian outside the centre. If
interpreted following both~\citet{Ge93} and~\citet{vdMF93}, this
suggests a radially biased orbital structure in Sculptor, Carina and
Sextans. Detailed dynamical modelling is required in order to quantify
the orbital structure, and to assess the effects of the stellar
density distribution as well as those of the unknown inclination.
Nonetheless, on a qualitative level, the sharply falling photometric
profile of the dSphs assures us that a significantly peaked velocity
dispersion can be robustly associated with a radial bias of the
orbits.  On the other hand, Fornax, shows line profiles that are
flat-topped at large radii, hence perhaps favouring some tangential
anisotropy. This suggests that it may have had a different recent
accretion history to the other dSphs. Support for this viewpoint is
also provided by its distinctive shell structures
\citep[e.g.,][]{Co05, Ol06, Co08}.

We also consider the angular behaviour of the asymmetric deviations
from Gaussianity. In Fornax, we find a systematic angular trend, that
we interpret as originating in a small level of intrinsic rotation.
In fact, we find that this trend is mirrored in systematic residuals
in the mean velocity with respect to the astrometrically determined
proper motion. This is consistent with a mild intrinsic rotation about
the minor axis, reaching about 1 kms$^{-1}$ in the radial range
covered by the kinematic sample.

Our methods for characterizing the shapes of line profiles in discrete
kinematic datasets are both powerful and adaptable.  In recent times,
the size and variety of such datasets has increased substantially,
with applications that range from the kinematics of the globular
cluster populations in relatively distant massive galaxies to
precision kinematics of giant stars in our own Galactic neighborhood.
We anticipate that our methods will find ready application to a rich
variety of datasets, and are actively pursuing further applications.

\section*{Acknowledgments}
It is a pleasure to thank Adriano Agnello, Giuseppe Bertin and Mike
Irwin for constructive discussions, as well as the anonymous
referee. We also thank Thomas de Boer for providing the kinematic
dataset pertaining to the Sculptor dSph. NA thanks STFC and the Isaac
Newton Trust for financial support.

\label{lastpage}

\appendix

\appendix

\section{Construction of asymmetric deviations}\label{asymmA}
\begin{figure}
\centering \includegraphics[width=\columnwidth]{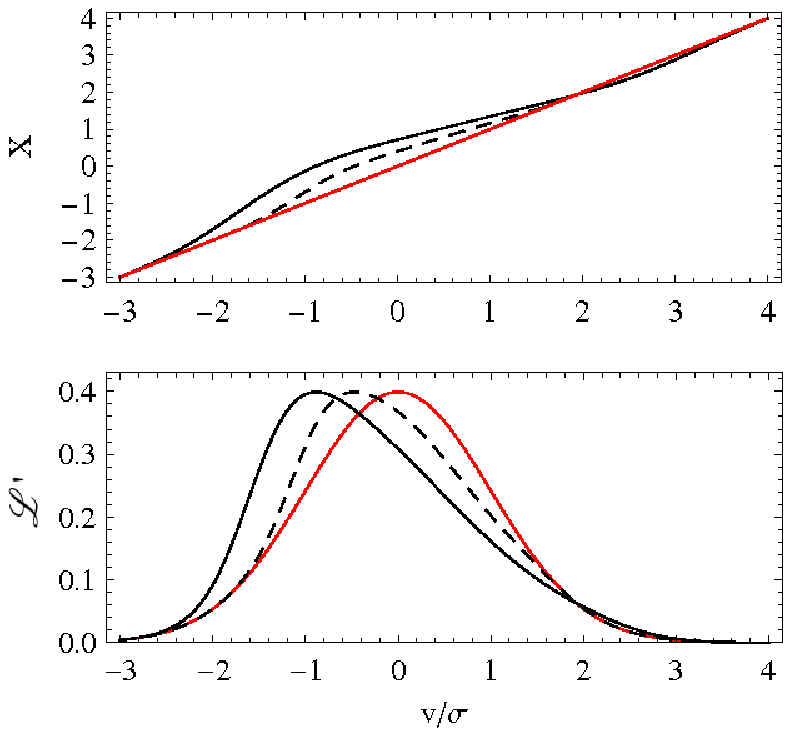}
\caption{The construction of asymmetric distributions. The upper panel
  displays the comparison between the identity function $\X(\s,\a=0;
  v)$ (in red) and two analogous functions (with $\s=0$) obtained for
  nonzero values of a (respectively $\a=0.3$ as a dashed line and
  $\a=0.6$ as a dashed line).  The lower panel illustrates, with the
  same graphical coding, the associated distributions $\LL'$,
  constructed as in eqn~(\ref{fasymm}).}
\label{asymmconstr}
\end{figure}

We considered several alternatives for the implementation of
asymmetric deviations starting with a simple parametrization
compatible with the general form
\begin{equation}
f=f_{\rm e}(\s)(1-\a f_{\rm o}(\s,\a))\ ,
\label{fevenodd}
\end{equation} 
where $f_{\rm e}$ and $f_{\rm o}$ are respectively an even and odd
function of one of the components of the tangential velocity
$\vec{v}_t$, for example $v_\theta$, and $a$ is the parameter for
asymmetric deviations. This approach has not been successful for at
least two reasons.

First, the asymmetric deviations produced by the functional
form~(\ref{fevenodd}) -- which has to satisfy the consistency
requirement $f\geq 0$ -- are too small for our purposes. It is
importnat to have a large template of deviations during the measuring
procedure.  Depending on the sample size and given the natural
accuracy limits we quantified in Sect~\ref{Nnoise}, even in the case
of an intrinsically symmetric distribution, distributions with a large
asymmetry ($h_3\gtrapprox 0.1$) are needed in order to assess a
reliable errorbar.  Second, the functional form~(\ref{fevenodd}) has
the significant limit of correlating symmetric and asymmetric
deviations. At fixed $s$, $f_{los}(v_{\parallel}=0)$ is in fact
invariant with respect to $a$, anh hence, asymmetric deviations come
together with a more spiky profile, which is not a desirable feature.

Our implementation of asymmetries, presented in eqn.~(\ref{fasymm}) is
able to overcome both difficulties. Fig.~\ref{asymmconstr} illustrates
the main ingredients of this approach.  The upper panel displays the
comparison between the identity function in red, representing the case
$\X(\s,\a=0; v)$, and two analogous functions obtained for $\s=0$ but
nonzero values of $a$.  In the lower panel we display the associated
distributions $\LL'$, in comparison with the Gaussian profile in
red. Our function $X$ in eqn~(\ref{Xdef}) is constructed in order to
comply with a series of requirements.
\begin{itemize} 
\item{$X$ is asymptotic to $v$ at both negative and positive extremes
  of the real axis. This is achieved by the structure $X-v\sim
  \exp(v^4)$.}
\item{The choice of the fourth power (rather than the second, for
  example) assures that asymmetric distributions are not significantly
  spikier than the associated symmetric distribution: $X$ and $v$ are
  almost parallel when $X\approx0$.}
\item{The dependences on $\a$ in the exponential term are required so
  to adapt the magnitude of the deviations of $X$ from $v$ to the size
  of the interval (as well as its position) where these deviations
  need to affect $\LL'$.}
\item{$X$ has to cross the identity function in order to guarantee a
  shallow decline on one of the wings of the asymmetric distribution,
  that in turn crosses the associated symmetric distribution. This is
  obtained by the linear term that multiplies the exponential one in
  $X$.}
\item{This crossing point is adjusted to both the shape of the
  symmetric distribution and to the magnitude of the asymmetric
  deviations by the dependences on $s$ and $a$ in the mentioned linear
  term.}
\item{Finally, the multiplication by $a/|a|$ ensures that the
  magnitude and shape of asymmetric deviations are identical (other
  than in direction) for positive and negative values of a.}
\end{itemize}

\end{document}